\documentclass[graphicx]{revtex4-1}

\begin{document}
\title{Isomorphism between the local Poincar\'e generalized translations group and the group of spacetime transformations $(\bigotimes \mathrm{LB1})^{4}$} 

\author{Alcides Garat}
\email[]{garat.alcides@gmail.com}
\affiliation{Former Professor at Universidad de la Rep\'{u}blica, Av. 18 de Julio 1824-1850, 11200  Montevideo, Uruguay.}
\date{\today}

\begin{abstract}
We will prove that there is a direct relationship between the Poincar\'e subgroup of translations, and the group of tetrad transformations LB1 introduced in a previous manuscript. LB1 is the group composed by $\mathrm{SO}(1,1)$ plus two kinds of discrete transformations. Translations have been extensively studied under the scope of gauge theories. By using the geometric structures built to prove this elementary result we will generalize it to the case of what we might call local translations.  A special case of the latter is the Bondi-Metzner-Sachs subgroup of supertranslations. In order to accomplish this goal and since the group of translations is four-dimensional we will prove first that it is isomorphic to $(\bigotimes \mathrm{LB1})^{4}$.
In order to prove this claim we will introduce a system of differential equations involving several kinds of fields. Abelian, non-Abelian, spinor, gravitational. These fields will constitute the structure needed to build local tetrads of a new kind that allow for the proof to be carried out with simplicity. Results already obtained involving similar but not equal tetrads will be useful in our constructions and demonstrations. Translations and generalized translations isomorphic to tensor products of LB1 groups are not trivial results. Because the LB1 group is composed by $\mathrm{SO}(1,1)$ and two discrete transformations.
\end{abstract}



\maketitle 

\section{Introduction}
\label{intro}

Through a series of manuscripts \cite{A,AC,LomCon,ATGU,AYM,gaugeinvmeth,A3,ASU3,ASUN} we have proved that there is a connection between local internal groups of transformations and local tetrad groups of transformations. Essentially the groups that we called LB1 and LB2. Using the very fields involved in different systems of differential equations and local duality transformations of second rank antisymmetric tensors we managed to build a new kind of tetrads \cite{MTW,CBDW,RW} with special properties. At every point in four-dimensional Lorentzian spacetimes we can define two orthogonal planes \cite{SCH}. The timelike tetrad vector and one spacelike define what we called blade or plane one. The other two spacelike vectors define an orthogonal blade or plane two. These new tetrads are relevant for several reasons. In the Abelian $\mathrm{U}(1)$ electromagnetic case in curved spacetimes or in Yang-Mills non-Abelian environments also in curved spacetimes of the kind $\mathrm{SU}(2) \times \mathrm{U}(1)$ or $\mathrm{SU}(3) \times \mathrm{SU}(2) \times \mathrm{U}(1)$ these tetrads diagonalize locally and covariantly the corresponding stress-energy tensors \cite{A,AC,LomCon,ATGU,AYM,gaugeinvmeth,A3,ASU3,ASUN}. Therefore the local orthogonal planes one or two are planes of diagonalization or symmetry. These tetrads have in their construction two fundamental components. These are what we call the tetrad skeletons and the tetrad gauge vectors. The tetrad skeletons are locally gauge invariant and are built out of extremal fields, which in turn are defined through local duality rotations, that will be defined in section \ref{newtetrads}. The gauge vectors are constructions that represent gauge per se and contain gauge in their structure. Local Lorentz transformations transform the tetrad extremal field-gauge vector structure into a similar field structure, property that we called structure invariance \cite{AYM}. The group $\mathrm{U}(1)$ was proven isomorphic to tetrad rotations inside blade two \cite{RG,GRSYMM}, that is, to the group $\mathrm{SO}(2)$ or LB2 as we called it since it represents local Lorentz tetrad spatial rotations inside blade two. That is, when we transform the local tetrad vectors that span blade two under the local gauge group $\mathrm{U}(1)$, the tetrad vectors rotate locally inside the original blade or plane that they generate.  Similarly, the group $\mathrm{U}(1)$ was proven isomorphic to tetrad boosts plus two different kinds of discrete transformations on blade one, that is, to the group LB1. The group LB1 is composed of boosts on plane one, $\mathrm{SO}(1,1)$ or Lorentz transformations on blade one and two different discrete tetrad transformations. One of the discrete transformations is not a Lorentz transformation, it is a vector interchange, flip or ``switch'' as we called it in reference \cite{A}. It is a reflection and therefore not a Lorentz transformation. In matrix form, two by two with zero diagonal and ones in both off-diagonal matrix components. The other discrete tetrad transformation is minus the identity in two by two matrix form, that is, a tetrad ``full inversion'' on plane one. Similarly in manuscript \cite{AYM} it was proven that the local internal gauge group $\mathrm{SU}(2)$ is isomorphic to both $(\bigotimes \mathrm{LB1})^{3}$ and $(\bigotimes \mathrm{LB2})^{3}$, independently. It is very important that we make a clarification on an issue that might give rise to confusion.
In these theorems for the Yang-Mills non-Abelian case like $\mathrm{SU}(2) \times \mathrm{U}(1)$ see references \cite{AYM,A3}  we are considering three copies of the same spacetime, and a different tetrad at the same point in each spacetime copy. These three tetrads have a mutually similar extremal field-gauge vector structure. They are normalized and a choice of gauge vector has been made. But they are not the same. They could be Lorentz transformed into each other, under non-trivial Lorentz spatial rotations, for instance. We know from manuscript \cite{AYM} that a local Lorentz transformation of a tetrad with an extremal field-gauge vector structure transforms into another tetrad with a similar extremal field-gauge vector structure, even though the skeletons will not be the same, of course. Therefore, we are considering three local tetrads at the same spacetime point which are not the same for different copies. This is what we mean by three LB1 or LB2 groups under tensor product. Now, from a practical point of view what we also mean by these theorems is that we are able to reconstruct the original local $\mathrm{SU}(2)$ gauge transformation by knowing for example, the boosts in the LB1 case, or the spatial rotations in the LB2 case, for the tetrad local transformations at the point under consideration. By knowing the local Lorentz transformation values for the three copies, and given all the fields, specially the three tetrads at the same point, we can reconstruct the local $\mathrm{SU}(2)$ transformation that gave rise either to three local LB1 transformations or independently to three local LB2 spatial transformations. Because the theorems already proved \cite{AYM,gaugeinvmeth,A3}, represent local isomorphisms between either three LB1 groups and $\mathrm{SU}(2)$ or independently three LB2 groups and $\mathrm{SU}(2)$. Similar for the case $\mathrm{SU}(3) \times \mathrm{SU}(2) \times \mathrm{U}(1)$ considering eight copies of the same spacetime, and a different tetrad at the same point in each spacetime copy \cite{ASU3}.

These results have paramount importance since it means that the hypotheses to the no-go theorems \cite{SWNG,LORNG,CMNG} are incorrect. Because of the following reasons. First, if all the local internal gauge transformations are isomorphic to local Lorentz tetrad transformations groups like LB1 and LB2 separately, then, it is not possible that the generators of internal groups commute with the generators of the Poincar\'e group. We read from reference \cite{CMNG} ``S (the scattering matrix) is said to be Lorentz-invariant if it possesses a symmetry group locally isomorphic to the Poincar\`{e} group P.\ldots A symmetry transformation is said to be an internal symmetry transformation if it commutes with P. This implies that it acts only on particle-type indices, and has no matrix elements between particles of different four-momentum or different spin. A group composed of such transformations is called an internal symmetry group''. The local electromagnetic gauge group of transformations $\mathrm{U}(1)$ has been proven to be isomorphic to local groups of tetrad transformations LB1 and LB2 on both the orthogonal planes one and two, see theorems {\bf V-VI-VII} in section \ref{sec:appIV}. The local orthogonal planes of Einstein-Maxwell stress-energy diagonalization. All vectors in these local orthogonal planes are eigenvectors of the stress-energy tensor. These local groups of transformations LB1 and LB2$=\mathrm{SO}(2)$ are composed of Lorentz transformations and even though the LB1 improper discrete reflection flip is not a Lorentz transformation, it is composed with this exception of spacetime Lorentz transformations, see reference \cite{A}. The spacetime flip is a discrete transformation given by $\Lambda^{o}_{\:\:o} = 0$, $\Lambda^{o}_{\:\:1} = 1$, $\Lambda^{1}_{\:\:o} = 1$, $\Lambda^{1}_{\:\:1} = 0$. We notice that this discrete transformation is not a Lorentz transformation because it is a reflection. Therefore the local Lorentz group of spacetime transformations cannot commute with LB1 or LB2 since Lorentz transformations on a local plane do not commute with Lorentz transformations on another local plane, necessarily. That is, the local internal groups of transformations do not necessarily commute with the local Lorentz transformations, because they are isomorphic to local groups of tetrad spacetime transformations on local orthogonal special planes, see section \ref{sec:appIV} for details and reference \cite{A}. The planes of diagonalization of the stress-energy tensor, which are unique. Similar results were proven for the Yang-Mills cases $\mathrm{SU}(2) \times \mathrm{U}(1)$ and $\mathrm{SU}(3) \times \mathrm{SU}(2) \times \mathrm{U}(1)$ geometrodynamics, see references \cite{AYM,gaugeinvmeth,A3,ASU3}. Second, because for every local element in $\mathrm{U}(1)$ there is an element in LB1 and another in LB2, $\mathrm{U}(1)$ is mapped into LB1 and LB2 independently, see section \ref{fusionkernel} in appendix \ref{sec:appIV}. All these construction elements bring about the possibility for microstructures to have associated spacetimes, since all their symmetries can be realized in four-dimensional Lorentzian spacetimes. These are the reasons why we set out to explore systems of differential equations and the tetrad sets that can be built using the available field structures coming out from them. We aim to produce through the tensor products of the groups LB1 and LB2 the unification of the standard model and general relativity on one hand, and on the other hand, grand group unification of all the local internal groups of the standard model and all the local spacetime groups of general relativity. As mentioned above, these tetrads have in their construction two essential components. We will define them properly in section \ref{newtetrads}, but we can anticipate that these are what we call the tetrad skeleton and the tetrad gauge vectors. The tetrad skeletons are locally gauge invariant. The gauge vectors are constructions that represent gauge by themselves and contain gauge fields in their structure. In this particular work it is our intention to define new tetrads, such that Poincar\'e transformations \cite{AOB} inside the tetrad gauge vectors allow us to make further group findings. Translations have been extensively studied under the scope of gauge theories, see reference \cite{69H}, specially chapter VI. Starting with author T. W. B. Kibble, see reference \cite{TWBK}, the tetrad field has been considered by many authors to be the gauge field for spacetime translations. The field that gauges translations, see the references in \cite{69H}, chapter VI. That is why we focus and relate translations and tetrad fields. In particular by using similar tetrads to the ones in paper \cite{AYM}, but with an important difference since now spinor fields will be available and we will substitute one of the electromagnetic tetrads inside the gauge vector for a current, all of this in section \ref{newtetrads}. In section \ref{tetradtransf} we will carry out these Poincar\'e transformations as a model of an elementary case, specially the translations by using the electromagnetic tetrad embedded inside the gauge vectors and finally prove two theorems about isomorphisms between the group of translations and the local group $(\bigotimes \mathrm{LB1})^{4}$, and also the group $(\bigotimes \mathrm{LB2})^{4}$. Then, we will proceed to the definition of local generalized translations and prove isomorphism group theorems in this more general case of our interest. A special case of the latter will be the Bondi-Metzner-Sachs subgroup of supertranslations. Throughout the paper we use the conventions of \cite{A,MW}. In particular we use a metric with sign conventions + - - -, and $f^{k}_{\mu\nu}$ are the geometrized Yang-Mills field components, $f^{k}_{\mu\nu}= (G^{1/2} / c^2) \: F^{k}_{\mu\nu}$.

\section{New Tetrads}
\label{newtetrads}

It is then appropriate to introduce at this point the differential equations we are considering in this work, the Einstein-Maxwell-Yang-Mills-Weyl differential equations,

\begin{eqnarray}
R_{\mu\nu} &=& T^{(ym)}_{\mu\nu} + T^{(Y)}_{\mu\nu} + \kappa_{sp}\:T^{(\mathrm{spinor})}_{\mu\nu} \label{eyme}\ ,\\
B^{Y\mu\nu}_{\:\:\quad;\nu} &=& \kappa_{a}\:j_{a}^{\mu} \label{emag1}\ ,\\
\ast B^{Y\mu\nu}_{\:\:\quad;\nu} &=& 0 \label{emag2}\ ,\\
f^{k\mu\nu}_{\:\:\quad\mid \nu} &=& \kappa_{na}\:j_{na}^{k\mu}\label{ymvfe1}\ ,\\
\ast f^{k\mu\nu}_{\:\:\quad\mid \nu} &=& 0  \label{ymvfe2}\ ,\\
\sigma^{\mu}\:{\cal{D}}_{\mu} \psi &=& 0 \ , \label{weylsu2}
\end{eqnarray}

We introduced in the system (\ref{eyme}-\ref{weylsu2}) as many fields as possible, so that in future works we can consider special cases to this more general system of differential equations. We introduce several elements of the system (\ref{eyme}-\ref{weylsu2}). For instance, the object ${\cal{D}}_{\mu} \psi = \partial_{\mu}\psi_{AM} - \Gamma_{\mu M}^{\:\:\:\:\:\:Q}\:\psi_{AQ} - \imath\:g\:A_{\mu A}^{\:\:\:\:\:\:Q}\:\psi_{QM} + \imath\:\frac{g^{'}}{2}\:B^{Y}_{\mu}\:\psi_{AM} = \partial_{\mu}\:\psi_{AM} + \frac{1}{2}\:\left( \sigma^{\beta\gamma} \right)_{M}^{\:\:\:\:Q}\:V_{\beta}^{\nu}\: V_{\gamma\nu;\mu}\:\psi_{AQ} - \imath \:g\:A_{\mu A}^{\:\:\:\:\:\:Q}\:\psi_{QM} + \imath\:\frac{g^{'}}{2}\:B^{Y}_{\mu}\:\psi_{AM}$, is the gauge covariant derivative, see sections \ref{sec:appI} and \ref{sec:appII} for details about the objects $\sigma^{\alpha}$, $\sigma^{\beta\gamma}$ and $\Gamma_{\mu M}^{\:\:\:\:\:\:Q}$. The gauge covariant derivatives of the three field strength internal components are given by $f_{k\mu\nu\mid\rho} = f_{k\mu\nu\, ; \, \rho} + g \: \epsilon_{klp}\: A_{l\rho}\:f_{p\mu\nu}$, where $\epsilon_{klp}$ is the completely skew-symmetric tensor in three dimensions with $\epsilon_{123} = 1$, and $g$ is the coupling constant. The symbol ``;'' stands for the usual covariant derivative associated with the metric tensor $g_{\mu\nu}$. We will call equation (\ref{weylsu2}) the wave equation, for covenience. The spinors $\psi$ belong to the tensor product space of two component $\mathrm{SU}(2)$ spinors, and two component $\mathrm{SL}(2,C)$ spinors. $\kappa_{sp}$, $\kappa_{a}$ and $\kappa_{na}$ are constants, the subindex $a$ stands for Abelian, and $na$ for non-Abelian. We will use $j_{a}^{\mu}$ without the $a$ subindex, for notational simplicity. $g$, $g^{'}$ are all standard coupling constants. In the object $\sigma^{\mu} = \sigma^{\alpha}\: V_{\alpha}^{\mu}$, the tetrad vectors $V_{\alpha}^{\:\:\mu}$ transform from a locally inertial coordinate system, into a general curvilinear coordinate system such that $\eta_{\alpha\beta} = g_{\mu\nu}\:V_{\alpha}^{\:\:\mu}\: V_{\beta}^{\:\:\nu}$, with $\eta_{\alpha\beta}$ the Minkowski metric tensor. In particular we use a metric with sign conventions +---, and $f^{k}_{\mu\nu}$ are the geometrized Yang-Mills field components, $f^{k}_{\mu\nu}= (G^{1/2} / c^2) \: F^{k}_{\mu\nu}$. Similar for the $\mathrm{U}(1)$ hypercharge geometrized field $B^{Y}_{\mu\nu} = \partial_{\mu}b_{\nu}^{Y} - \partial_{\nu}b_{\mu}^{Y}$, where $ b_{\mu}^{Y} = (G^{1/2} / c^2) \: B_{\mu}^{Y}$, see papers \cite{DG,HM,GR}. Now, with all these elements we proceed to introduce the tetrads that we will employ in our local group isomorphisms. The new tetrads have two basic components. The skeleton, and the gauge vector. For instance, and as a model for the present work we present once again the electromagnetic tetrad \cite{A},

\begin{eqnarray}
U^{\rho} &=& \xi^{\rho\lambda}\:\xi_{\tau\lambda}\:A^{\tau} \:
/ \: (\: \sqrt{-Q/2} \: \sqrt{A_{\mu} \ \xi^{\mu\sigma} \
\xi_{\nu\sigma} \ A^{\nu}}\:) \ ,\label{U}\\
V^{\rho} &=& \xi^{\rho\lambda}\:A_{\lambda} \:
/ \: (\:\sqrt{A_{\mu} \ \xi^{\mu\sigma} \
\xi_{\nu\sigma} \ A^{\nu}}\:) \ ,\label{V}\\
Z^{\rho} &=& \ast \xi^{\rho\lambda} \: \ast A_{\lambda} \:
/ \: (\:\sqrt{\ast A_{\mu}  \ast \xi^{\mu\sigma}
\ast \xi_{\nu\sigma}  \ast A^{\nu}}\:)\ ,
\label{Z}\\
W^{\rho} &=& \ast \xi^{\rho\lambda}\: \ast \xi_{\tau\lambda}
\:\ast A^{\tau} \: / \: (\:\sqrt{-Q/2} \: \sqrt{\ast A_{\mu}
\ast \xi^{\mu\sigma} \ast \xi_{\nu\sigma} \ast A^{\nu}}\:) \ .
\label{W}
\end{eqnarray}

This tetrad was built for a system of Einstein-Maxwell differential equations of the kind,

\begin{eqnarray}
f^{\mu\nu}_{\quad;\nu} &=& 0 \ ,\label{EM1}\\
\ast f^{\mu\nu}_{\quad;\nu} &=& 0 \ ,\label{EM2}\\
R_{\mu\nu} &=& f_{\mu\lambda}\:\:f_{\nu}^{\:\:\:\lambda}
+ \ast f_{\mu\lambda}\:\ast f_{\nu}^{\:\:\:\lambda}\ , \label{EM3}
\end{eqnarray}

$f_{\mu\nu}$ is the geometrized electromagnetic field. We designate these four local vectors (\ref{U}-\ref{W}) by the generic expression $E_{\alpha}^{\:\:\rho}$. In this tetrad, in vector $U^{\rho}$ the skeleton would be $\xi^{\rho\lambda}\:\xi_{\tau\lambda}$, and $A^{\tau}$ the electromagnetic gauge vector, where $\xi_{\tau\lambda} = \cos\alpha\:f_{\tau\lambda} - \sin\alpha\:\ast f_{\tau\lambda}$ was defined as the extremal field \cite{A}, fulfilling the equation or imposed condition $\xi_{\mu\nu} \ast \xi^{\mu\nu}= 0$, and $Q = \xi_{\mu\nu}\:\xi^{\mu\nu}$ is assumed not to be zero because the electromagnetic field $f_{\tau\lambda}$ is non-null. Non-null we clarify means basically that $f_{\mu\nu}\:f^{\mu\nu}\neq0$ and $\ast f_{\mu\nu}\:f^{\mu\nu}\neq0$. In turn and by definitions these last equations imply that $\xi_{\mu\nu}\:\xi^{\mu\nu}\neq0$. As antisymmetric fields in a four-dimensional Lorentzian spacetime, the extremal fields also verify the identity,

\begin{eqnarray}
\xi_{\mu\alpha}\:\xi^{\nu\alpha} -
\ast \xi_{\mu\alpha}\: \ast \xi^{\nu\alpha} &=& \frac{1}{2}
\: \delta_{\mu}^{\:\:\:\nu}\ Q \ .\label{i1}
\end{eqnarray}

When we use the general identity,

\begin{eqnarray}
A_{\mu\alpha}\:B^{\nu\alpha} -
\ast B_{\mu\alpha}\: \ast A^{\nu\alpha} &=& \frac{1}{2}
\: \delta_{\mu}^{\:\:\:\nu}\: A_{\alpha\beta}\:B^{\alpha\beta}  \ ,\label{2ig}
\end{eqnarray}

which is valid for every pair of antisymmetric tensors in a four-dimensional Lorentzian spacetime \cite{MW}, and apply it to the case
$A_{\mu\alpha} = \xi_{\mu\alpha}$ and $B^{\nu\alpha} = \ast \xi^{\nu\alpha}$, it yields an equivalent condition to $\xi_{\mu\nu} \ast \xi^{\mu\nu}= 0$,

\begin{eqnarray}
\xi_{\alpha\mu}\:\ast \xi^{\mu\nu} &=& 0\ .\label{rceq}
\end{eqnarray}

By means of this general identity (\ref{2ig}) it is simple to prove that the condition $\xi_{\mu\nu} \ast \xi^{\mu\nu}= 0$ is equivalent to $\xi_{\mu\nu} \ast \xi^{\mu\rho}= 0$. The local scalar complexion arises as $\tan(2\alpha) = - f_{\mu\nu}\:\ast f^{\mu\nu} / f_{\lambda\rho}\:f^{\lambda\rho}$, from the condition $\xi_{\mu\nu} \ast \xi^{\mu\nu}= 0$ which is in turn equivalent to equation $\xi_{\mu\nu} \ast \xi^{\mu\rho}= 0$. In vector $Z^{\rho}$ the skeleton is $\ast \xi^{\rho\lambda}$, and $\ast A_{\lambda}$ the second electromagnetic gauge vector \cite{CF}. We remind ourselves that the dual to the tensor $\xi_{\mu\nu}$ is $\ast \xi_{\mu\nu}={1 \over 2}\:\epsilon_{\mu\nu\sigma\tau}\:\xi^{\sigma\tau}$. Because of the Maxwell equations for the Abelian case, we can write $f_{\mu\nu} = A_{\nu ;\mu} - A_{\mu ;\nu}$ and $\ast f_{\mu\nu} = \ast A_{\nu ;\mu} - \ast A_{\mu ;\nu}$. The star in $\ast A_{\lambda}$ is not the Hodge map, is just notation, meaning that $\ast A_{\nu ;\mu} = (\ast A_{\nu})_{;\mu}$. Because of the two Maxwell equations within the coupled system of equations (\ref{EM1}-\ref{EM2}), two vector potentials are available $A_{\nu}$ and $\ast A_{\nu}$, see reference \cite{CF}. In this manuscript, and with the purpose of proving group results in the next section, we will consider the tetrad constructed similarly to the electromagnetic tetrad, but with the Hypercharge potential instead as our fundamental tool, even though we could also consider for example, the $\mathrm{SU}(2)$ tetrad, see paper \cite{AYM}. In order to be succinct we just would like to mention the main steps in the construction of the tetrad adapted to our new model of differential equations, according to their field content. We can define then the hypercharge geometrized field $B^{Y}_{\mu\nu} = \partial_{\mu}b_{\nu}^{Y} - \partial_{\nu}b_{\mu}^{Y}$. Applying to this antisymmetric Hypercharge field all the techniques developed in paper \cite{A} we can define a similar extremal Hypercharge field $\xi^{Y}_{\tau\lambda} = \cos\alpha^{Y}\:B^{Y}_{\tau\lambda} - \sin\alpha^{Y}\:\ast B^{Y}_{\tau\lambda}$, satisfying $\xi^{Y}_{\mu\nu} \ast \xi^{Y \mu\rho}= 0$, and finally a local complexion $\alpha^{Y}$, and a tetrad ($U^{Y\rho},V^{Y\rho},Z^{Y\rho},W^{Y\rho}$) built in a analogous way to the electromagnetic tetrad. Knowing that the extremal field is $\xi^{Y}_{\tau\lambda}$ with identical properties to the electromagnetic case, we proceed to suppress the $Y$ in all quantities in order to keep the notation simple. In tetrad (\ref{U}-\ref{W}) the tetrad gauge vectors are $X^{\nu} = A^{\nu}$ and $Y^{\nu} = \ast A^{\nu}$, see paper \cite{CF}. For a general Hypercharge tetrad,

\begin{eqnarray}
U^{\rho} &=& \xi^{\rho\lambda}\:\xi_{\tau\lambda}\:X^{\tau} \:
/ \: (\: \sqrt{-Q/2} \: \sqrt{X_{\mu} \ \xi^{\mu\sigma} \
\xi_{\nu\sigma} \ X^{\nu}}\:) \ ,\label{UG}\\
V^{\rho} &=& \xi^{\rho\lambda}\:X_{\lambda} \:
/ \: (\:\sqrt{X_{\mu} \ \xi^{\mu\sigma} \
\xi_{\nu\sigma} \ X^{\nu}}\:) \ ,\label{VG}\\
Z^{\rho} &=& \ast \xi^{\rho\lambda} \: \ast Y_{\lambda} \:
/ \: (\:\sqrt{\ast Y_{\mu}  \ast \xi^{\mu\sigma}
\ast \xi_{\nu\sigma}  \ast Y^{\nu}}\:)\ ,
\label{ZG}\\
W^{\rho} &=& \ast \xi^{\rho\lambda}\: \ast \xi_{\tau\lambda}
\:\ast Y^{\tau} \: / \: (\:\sqrt{-Q/2} \: \sqrt{\ast Y_{\mu}
\ast \xi^{\mu\sigma} \ast \xi_{\nu\sigma} \ast Y^{\nu}}\:) \ ,
\label{WG}
\end{eqnarray}

we could choose gauge vectors adapted to our proofs. Let us introduce them,

\begin{eqnarray}
X^{\sigma} = Y^{\sigma} = \mathrm{Tr}[\Sigma^{\alpha\beta}\:J_{\alpha}\: S_{\beta}^{\:\:\lambda}\:\ast \epsilon^{\sigma}_{\:\:\:\lambda;\tau}\:A^{\tau}] \ . \label{gaugevectors}
\end{eqnarray}

Let us call the Hypercharge tetrad vectors (\ref{UG}-\ref{WG}) by the generic name $B_{\alpha}^{Y\:\mu}$, or dropping the $Y$, just $B_{\alpha}^{\mu}$. In expression (\ref{gaugevectors}) $J^{\alpha} = \psi^{\dagger}\sigma^{\alpha}\psi$ is the current associated to the spinor field \cite{GR2}, see section \ref{sec:appII}. $S_{\beta}^{\:\:\lambda}$ are the $\mathrm{SU}(2)$ tetrads introduced in paper \cite{AYM} and section \ref{sec:appIII}. Let us remember the expression for the $\mathrm{SU}(2)$ tetrads,

\begin{eqnarray}
S_{o}^{\mu} &=& \epsilon^{\mu\lambda}\:\epsilon_{\rho\lambda}\:X^{\rho} \:
/ \: (\: \sqrt{-Q_{ym}/2} \: \sqrt{X_{\mu} \ \epsilon^{\mu\sigma} \
\epsilon_{\nu\sigma} \ X^{\nu}}\:) \ ,\label{S1}\\
S_{1}^{\mu} &=& \epsilon^{\mu\lambda} \: X_{\lambda} \:
/ \: (\:\sqrt{X_{\mu} \ \epsilon^{\mu\sigma} \
\epsilon_{\nu\sigma} \ X^{\nu}}\:) \ ,\label{S2}\\
S_{2}^{\mu} &=& \ast \epsilon^{\mu\lambda} \: Y_{\lambda} \:
/ \: (\:\sqrt{Y_{\mu}  \ast \epsilon^{\mu\sigma}
\ast \epsilon_{\nu\sigma} Y^{\nu}}\:) \ ,\label{S3}\\
S_{3}^{\mu} &=& \ast \epsilon^{\mu\lambda}\: \ast \epsilon_{\rho\lambda} \: Y^{\rho}/ \: (\:\sqrt{-Q_{ym}/2} \: \sqrt{Y_{\mu}
\ast \epsilon^{\mu\sigma} \ast \epsilon_{\nu\sigma} Y^{\nu}}\:) \ ,\label{S4}
\end{eqnarray}

where $Q_{ym} = \epsilon_{\mu\nu}\:\epsilon^{\mu\nu}$ is assumed not to be zero. This new kind of local $\mathrm{SU}(2)$ gauge invariant extremal tensor $\epsilon_{\mu\nu}$, allows for the construction of this new tetrad $S_{\beta}^{\:\:\mu}$, see section \ref{sec:appIII}. The skeletons in (\ref{S1}-\ref{S4}) are gauge invariant under local $\mathrm{SU}(2) \times \mathrm{U}(1)$ gauge transformations, see papers \cite{A,AC,LomCon,AYM}. This property guarantees that the vectors (\ref{S1}-\ref{S4}) under local $\mathrm{U}(1)$ or $\mathrm{SU}(2)$ gauge transformations will not leave their original planes or blades, keeping therefore the metric tensor explicitly invariant. Let us remind ourselves that the local $\mathrm{SU}(2)$ tetrad is invariant under $\mathrm{U}(1)$ local gauge transformations, the whole tetrad, not only the skeletons. The gauge vectors chosen in paper \cite{AYM} for the $\mathrm{SU}(2)$ tetrad were chosen such that they are invariant under local $\mathrm{U}(1)$ gauge transformations, $X^{\sigma} = Y^{\sigma} = Tr[\Sigma^{\alpha\beta}\:B_{\alpha}^{\:\:\rho}\: B_{\beta}^{\:\:\lambda}\:\ast \xi_{\rho}^{\:\:\sigma}\:\ast \xi_{\lambda\tau}\:A^{\tau}]$. This is because the object $B_{[\alpha}^{\:\:\rho}\: B_{\beta]}^{\:\:\lambda}\:\ast \xi_{\rho}^{\:\:\sigma}\:\ast \xi_{\lambda\tau}=B_{[2}^{\:\:\rho}\: B_{3]}^{\:\:\lambda}\:\ast \xi_{\rho}^{\:\:\sigma}\:\ast \xi_{\lambda\tau}$ is invariant under $\mathrm{U}(1)$ local gauge transformations where $\xi_{\lambda\tau}$ is in this case the Hypercharge extremal field $\xi^{Y}_{\tau\lambda}$. We replaced the electromagnetic tetrad $E_{\alpha}^{\:\:\rho}$ used in paper \cite{AYM} for the Hypercharge tetrad $B_{\alpha}^{\mu}$ with similar properties. These are the gauge vectors for the $\mathrm{SU}(2)$ tetrad vectors (\ref{S1}-\ref{S4}).

\section{Poincar\'e tetrad transformations}
\label{tetradtransf}

It is the purpose of this work to probe the behavior of the Hypercharge tetrads (\ref{UG}-\ref{WG}) under a local ``translation'' of the $\mathrm{SU}(2)$ tetrad inside the new gauge vector (\ref{gaugevectors}). We are not translating the whole tetrad structure, just the $\mathrm{SU}(2)$ tetrad inside the new gauge vector (\ref{gaugevectors}). To this purpose we introduce the local transformation,

\begin{eqnarray}
S_{\beta}^{\:\:\mu} \rightarrow S_{\beta}^{\:\:\mu} + \varepsilon \: T_{\beta}^{\:\:\mu} \ , \label{TRANSL}
\end{eqnarray}

where $\varepsilon$ is a small parameter, and the vectors $T_{\beta}^{\:\:\mu}$ are given by,

\begin{eqnarray}
T_{\beta}^{\:\:\mu} = \mathfrak{a}^{\nu}\:\partial_{\nu}\:S_{\beta}^{\:\:\mu}\ . \label{TVECTOR}
\end{eqnarray}

The vectors $\mathfrak{a}^{\nu}$ for all $\beta = 0\ldots3$, are global constant arbitrary vectors. These components would be arbitrary and general, taking any possible real values for all four components of $\mathfrak{a}^{\nu}$. We would like to take this opportunity to discuss in more depth what is the general rationale of these kind of translations (\ref{TRANSL}-\ref{TVECTOR}). The first interrogant that comes to mind is why are we only translating the non-Abelian tetrad $S_{\beta}^{\:\:\nu}$ inside the Hypercharge gauge vectors (\ref{gaugevectors}) and not translating anything else. Because if we are implementing a translation $x^{\mu} \rightarrow x^{\mu}+a^{\mu}$ then, why are we not translating any other field in the tetrad structure of (\ref{UG}-\ref{VG}) for example. The answer is the following: we have freedom to choose the gauge vectors as we please, for instance the gauge vector $X^{\sigma}$ and then to study the transformation of $\xi^{\rho\lambda}\:\xi_{\tau\lambda}\:X^{\tau}$ or its normalized version (\ref{UG}) and similar for (\ref{VG}). This is an enormous advantage of our new tetrad methods. The actual translation in the Yang-Mills tetrad $S_{\beta}^{\:\:\nu}$ inside the Hypercharge gauge vectors (\ref{gaugevectors}) is nothing but another choice for the gauge vector $X^{\sigma}$ in (\ref{UG}). On top of all that, the Yang-Mills tetrad vectors (\ref{S1}-\ref{S4}) $S_{\beta}^{\:\:\nu}$ were built with independence of the electromagnetic tetrad vectors (\ref{U}-\ref{W}) and also with independence of the Hypercharge tetrad vectors (\ref{UG}-\ref{WG}) and we built them to be $\mathrm{U}(1)$ gauge invariant on purpose, both the skeletons and the gauge vectors associated to the Yang-Mills tetrad (\ref{S1}-\ref{S4}). Therefore when we make a ``translation'' transformation on the Yang-Mills tetrad embedded inside the Hypercharge gauge vector (\ref{gaugevectors}) we are just making another different choice for the Hypercharge gauge vectors of (\ref{UG}-\ref{VG}) using our previous choice (\ref{gaugevectors}) as a departure point. But simultaneously this is a true translation $x^{\mu} \rightarrow x^{\mu}+\varepsilon \:a^{\mu}$ of just the Yang-Mills tetrad $S_{\beta}^{\:\:\nu}$. This way we have enough gauge latitude or gauge freedom inside the gauge vectors in order to study group isomorphisms. This is the rationale behind all of these tetrad group transformations and gauge vector choices.

We proceed then to study the transformation of $\xi^{\rho\lambda}\:\xi_{\tau\lambda}\:X^{\tau}$ in (\ref{UG}) with $X^{\sigma}$  given by (\ref{gaugevectors}) under the transformation (\ref{TRANSL}-\ref{TVECTOR}),

\begin{eqnarray}
\xi^{\rho\lambda}\:\xi_{\tau\lambda}\: \mathrm{Tr}[\Sigma^{\alpha\beta}\:J_{\alpha}\: S_{\beta}^{\:\:\nu}\:\ast \epsilon^{\tau}_{\:\:\:\nu;\mu}\:A^{\mu}] \rightarrow \xi^{\rho\lambda}\:\xi_{\tau\lambda}\:\left( \mathrm{Tr}[\Sigma^{\alpha\beta}\:J_{\alpha}\: S_{\beta}^{\:\:\nu}\:\ast \epsilon^{\tau}_{\:\:\:\nu;\mu}\:A^{\mu}]  \right.  \nonumber \\ + \:\varepsilon\: \left.  \mathrm{Tr}[\Sigma^{\alpha\beta}\:J_{\alpha}\: T_{\beta}^{\:\:\nu}\:\ast \epsilon^{\tau}_{\:\:\:\nu;\mu}\:A^{\mu}] \right) \ . \label{TRSo}
\end{eqnarray}

This expression can be subsequently written explicitly as,

\begin{eqnarray}
\xi^{\rho\lambda}\:\xi_{\tau\lambda}\: \mathrm{Tr}[\Sigma^{\alpha\beta}\:J_{\alpha}\: S_{\beta}^{\:\:\nu}\:\ast \epsilon^{\tau}_{\:\:\:\nu;\mu}\:A^{\mu}] \rightarrow \xi^{\rho\lambda}\:\xi_{\tau\lambda} && \:\left( \mathrm{Tr}[\Sigma^{\alpha\beta}\:J_{\alpha}\: S_{\beta}^{\:\:\nu}\:\ast \epsilon^{\tau}_{\:\:\:\nu;\mu}\:A^{\mu}]  \right.  \nonumber \\ && \: + \varepsilon \left.  \mathrm{Tr}[\Sigma^{\alpha \beta}\:J_{\alpha}\: \mathfrak{a}^{0}\:\partial_{0}\:S_{\beta}^{\:\:\nu}\:\ast \epsilon^{\tau}_{\:\:\:\nu;\mu}\:A^{\mu}]  \right.  \nonumber \\ && \: + \varepsilon \left.  \mathrm{Tr}[\Sigma^{\alpha \beta}\:J_{\alpha}\: \mathfrak{a}^{1}\:\partial_{1}\:S_{\beta}^{\:\:\nu}\:\ast \epsilon^{\tau}_{\:\:\:\nu;\mu}\:A^{\mu}]  \right.  \nonumber \\
&& + \: \varepsilon \left.  \mathrm{Tr}[\Sigma^{\alpha \beta}\:J_{\alpha}\: \mathfrak{a}^{2}\:\partial_{2}\:S_{\beta}^{\:\:\nu}\:\ast \epsilon^{\tau}_{\:\:\:\nu;\mu}\:A^{\mu}]  \right.  \nonumber \\
&& + \:\varepsilon \left.  \mathrm{Tr}[\Sigma^{\alpha \beta}\:J_{\alpha}\: \mathfrak{a}^{3}\:\partial_{3}\:S_{\beta}^{\:\:\nu}\:\ast \epsilon^{\tau}_{\:\:\:\nu;\mu}\:A^{\mu}] \right)\ . \label{TRSoexpanded2}
\end{eqnarray}

Remembering all our previous analysis on this transformation structure, specially from papers \cite{A,AC,LomCon,ATGU,AYM,gaugeinvmeth,A3,ASU3,ASUN} in their respective ``gauge geometry'' sections, we conclude the following. These local translations, generated by (\ref{TRANSL}-\ref{TVECTOR}) have all the necessary group properties. If all the $\mathfrak{a}^{\delta} = 0,\:\: \mbox{for} \:\: \delta:o \ldots 3$, then we produce the identity in LB1. Every local translation has an inverse in LB1 generated by $-\mathfrak{a}^{\delta},\:\: \mbox{for} \:\: \delta:o \ldots 3$. They are obviously associative since they are locally additive. The demonstration of the injectivity and the surjectivity of these local transformations would follow the same lines as the analogous proof in paper \cite{A}. We are not repeating these results in this paper because it would be redundant. All these LB1 local transformations independent from each other, since the terms in expression (\ref{TRSoexpanded2}) do not mix the $\mathfrak{a}^{\delta},\:\: \mbox{for} \:\: \delta:o \ldots 3$. A complete similar analysis can be done for the vector (\ref{VG}) on the local blade one, and subsequently for vectors (\ref{ZG}-\ref{WG}) on the local blade two. These means in turn that local translations of the local $\mathrm{SU}(2)$ tetrads $S_{\beta}^{\:\:\nu}$ inside the Hypercharge gauge vectors (\ref{gaugevectors}) generate local LB1 transformations of the normalized Hypercharge tetrad vectors (\ref{UG}-\ref{WG}), Abelian among them, and since we have four arbitrary components for the vector $\mathfrak{a}^{\mu}$, the terms in expression (\ref{TRSoexpanded2}) are additive and do not mix components. Based on similar lines of analysis to the one carried out in manuscript \cite{A}, the image of this mapping (\ref{TRANSL}-\ref{TVECTOR}) is not a subgroup of LB1. As in paper \cite{AYM} we proceed with a replica analysis of all the elements introduced in sections \ref{intro}, \ref{newtetrads} and \ref{tetradtransf}.
In accordance to the same guidelines as in papers \cite{A,AC,LomCon,ATGU,AYM} we conclude that,

\newtheorem {guesslb1} {Theorem}
\begin{guesslb1}
The mapping between the subgroup of Poincar\'e translations and the tensor product of the four local groups of $\mathrm{LB1}$ tetrad transformations is isomorphic. \end{guesslb1}

Following analogously the reasoning laid out in \cite{A,AC,LomCon,AYM}, in addition to the ideas above, we can also state,

\begin{guesslb1}
The mapping between the subgroup of Poincar\'e translations and the tensor product of the four local groups of $\mathrm{LB2}$ tetrad transformations is isomorphic. \end{guesslb1}

We observe additionally the ensuing remarkable property. Should we introduce the local generalized transformations with $c^{\nu}$ a vector with general local components in equations (\ref{TRANSL}-\ref{TVECTOR}),

\begin{eqnarray}
T_{\beta}^{\:\:\mu} = \mathfrak{c}^{\:\:\nu}\:\partial_{\nu}\:S_{\beta}^{\:\:\mu}\ . \label{TVECTORGEN}
\end{eqnarray}

and using the transformation (\ref{TRSoexpanded2}) we can analyze the following. If all the local components $c^{\nu} = 0,\:\: \mbox{for} \:\: \nu:o \ldots 3$, then we produce the identity in LB1. Every local generalized translation has an inverse in LB1 generated by $-c^{\nu},\:\: \mbox{for} \:\: \nu:o \ldots 3$. They are obviously associative since they are locally additive. The demonstration of the injectivity and the surjectivity of these local transformations would follow the same lines as the analogous proof in paper \cite{A}, see section \ref {injsur} in appendix \ref{sec:appIV} for more details. Once again, we are not repeating these results in this paper because it would be redundant even though we are presenting some elements of analysis in section \ref{sec:appIV}. All these four LB1 local transformations in equation (\ref{TVECTORGEN}) independent from each other, since the terms in expression (\ref{TRSoexpanded2}) do not mix the $c^{\nu},\:\: \mbox{for} \:\: \nu:o \ldots 3$. A complete analysis can be done for the vectors (\ref{UG}-\ref{VG}) on the local blade one, and subsequently for vectors (\ref{ZG}-\ref{WG}) on the local blade two. These means in turn that local translations of the local $\mathrm{SU}(2)$ tetrads $S_{\beta}^{\:\:\nu}$ inside the Hypercharge gauge vectors (\ref{gaugevectors}) generate local LB1 transformations of the normalized Hypercharge tetrad vectors (\ref{UG}-\ref{WG}), Abelian among them, as expected. And again based on similar lines of analysis to the ones carried out in manuscripts \cite{A,AC,LomCon,ATGU,AYM}, the image of this mapping (\ref{TRANSL}-\ref{TVECTOR}) and (\ref{TVECTORGEN}) is not a subgroup of LB1, see section \ref{sec:appIV}. We conclude the following.

\begin{guesslb1}
The mapping between the local group of transformations (\ref{TVECTORGEN}) and the tensor product of the four local groups of $\mathrm{LB1}$ tetrad transformations is isomorphic. \end{guesslb1}

By similar reasoning to the one laid out in \cite{A,AC,LomCon,ATGU,AYM}, in addition to all the ideas above, we can also state,

\begin{guesslb1}
The mapping between the local group of transformations (\ref{TVECTORGEN}) and the tensor product of the four local groups of $\mathrm{LB2}$ tetrad transformations is isomorphic. \end{guesslb1}

We additionally can observe that the local group of transformations (\ref{TVECTORGEN}) for the particular cases of asymptotically flat curved spacetimes coincides with the Bondi-Metzner-Sachs \cite{AE,JG} subgroup of supertranslations.


\section{Conclusions}
\label{conclusions}

We have proven that the local group of Poincar\'e translations is isomorphic to the tensor product $(\bigotimes \mathrm{LB1})^{4}$. We also considered local generalized translations and proved isomorphism group theorems in this more general case of our interest. A special case of the latter is the Bondi-Metzner-Sachs subgroup of supertranslations. It is relevant that we make one more time a clarification on an issue that might give rise to confusion.
In theorems (1-2) we are considering four copies of the same spacetime, and a different tetrad at the same point in each spacetime copy. These four tetrads have a mutually similar extremal field-gauge vector structure. They are normalized and a choice of gauge vector has been made. But they are not the same. They could be Lorentz transformed into each other, under non-trivial Lorentz spatial rotations, for example. We know from manuscript \cite{AYM} that a local Lorentz transformation of a tetrad with an extremal field-gauge vector structure transforms into another tetrad with a similar extremal field-gauge vector structure, not the same local extremal field. Therefore, we are considering four local tetrads at the same spacetime point which are not the same for different copies. The local planes one and two will be tilted with respect to each other. This is what we mean by four LB1 or LB2 groups under tensor product. Now, from a practical point of view what we also mean by these theorems is that we are able to reconstruct the original constant translation (\ref{TRANSL}-\ref{TVECTOR}) or the local generalized translation (\ref{TVECTORGEN}) by knowing for example, the boosts in the LB1 case, or the spatial rotations in the LB2 case, for the tetrad local transformations at the point under consideration. By knowing the local Lorentz transformation values for the four copies, and given all the fields, specially the four tetrads at the same point, we can reconstruct the original constant translation (\ref{TRANSL}-\ref{TVECTOR}) or the local generalized translation (\ref{TVECTORGEN}) that gave rise either to four local LB1 transformations or independently to four local LB2 spatial transformations. Because the theorems proved, represent local isomorphisms between either four LB1 groups and the constant or local translations or independently four LB2 groups and the constant or local translations. This result adds to the set of theorems already proved on isomorphism between local groups of gauge transformations and tensor products either of LB1 or LB2 local groups. The fact that local translations are isomorphic to local tensor products of LB1 groups is not trivial, or evident. Because the LB1 group is composed by $\mathrm{SO}(1,1)$ and two discrete transformations. It becomes a straightforward fact through the use of the new tetrads with special skeletons and specially designed gauge vectors.
It is through these field architectures that we can effectively prove these results. Translations have been extensively studied within the realm of gauge theories, see reference \cite{69H}, specially chapter VI. Starting with the pioneering work of D. W. Sciama \cite{DWS} and T. W. B. Kibble, see reference \cite{TWBK}, the tetrad field has been considered as the gauge field for spacetime translations. The field that gauges translations, see the references in \cite{69H}, chapter VI. Hence the importance of these new tetrads. It is possible to establish a parallelism between the local transformation properties in the case of the electromagnetic tetrads under $\mathrm{U}(1)$ transformations \cite{A} on one hand, and the transformation of Hypercharge tetrads under local translations of the $\mathrm{SU}(2)$ tetrads nested inside their gauge vectors on the other hand. In the electromagnetic case the local tetrad gauge transformations through a gauge vector chosen to be the electromagnetic potential, on blade one for instance, generate LB1 local transformations. In the Hypercharge case studied in this paper the translations generate tensor products of LB1 local transformations. Therefore, we can see as an example that if we make three of the local scalars that generate translations to be zero, then the group of one dimensional translations that is left, acts as a local electromagnetic gauge transformation. In the $\mathrm{SU}(2)$ case, and we are not talking about the tetrads in this paper, but of the tetrads in manuscript \cite{AYM}, the main difference with $\mathrm{U}(1)$ let us say on blade one, is that the electromagnetic tetrads nested within its gauge vector undergo an additional local spatial rotation under locally generated $\mathrm{SU}(2)$ gauge transformations. Which are local Poincar\'e transformations as well. All these results unexpected under standard Riemannian geometry, or standard model physics, but natural through the use of the new tetrads. We aim to produce through the tensor products of the groups LB1 and LB2 the unification of the standard model and general relativity on one hand, and on the other hand grand group unification of all the local internal groups of the standard model and all the local spacetime groups of general relativity. As a final comment we would like to say that through the use of perturbative treatments we aim simultaneously to make a connection with quantum field theories as in references \cite{dsmg,dsmg1,DSBYM}. We quote from \cite{CIJB} ``It should be admitted from the outset, that the subject of quantum gravity is exceptionally difficult and problematic, both in regard to mathematical and to conceptual issues. Furthermore, these problems are compounded by the almost complete lack of any unequivocal data to guide attempts to construct such a theory. In this sense, twentieth-century physics is a victim of its own success: the empirical success of general relativity and quantum theory in their present forms means that we lack data bearing on how we might reconcile them (or more generally, replace them)''.

\section{Appendix I}
\label{sec:appI}

We are introducing in this first appendix the variables that we will need in order to construct the tetrads \cite{MC,PR}. Either the ``skeleton'' of the tetrad, or the gauge vector fields $X^{\mu}$ and $Y^{\mu}$. These are not all variables in terms of which the differential equations (\ref{eyme}-\ref{weylsu2}) are written, but they do exist and we are using them. We follow the notation in \cite{MC}. We start with the Hermitian matrices $\sigma^{\mu}_{AB'}$ introduced by Infeld and van der Waerden,

\begin{eqnarray}
\sigma^{\mu}_{AB'}\:\sigma^{\nu AB'} = g^{\mu\nu} \ .\label{metricupcurv}
\end{eqnarray}

Roman capital indices are the spinor indices taking values $0$ and $1$, while the primed Roman capital indices refer to the complex conjugate taking values $0'$ and $1'$. A detailed discussion is included in reference \cite{MC} chapter eight. Therefore we summarize the main relations and expressions involved in our work.

\begin{eqnarray}
\sigma^{\mu}_{AB'}\:\sigma^{\nu}_{CD'}\:g_{\mu\nu } = g_{AB'CD'} \ . \label{metricdownspin}
\end{eqnarray}

\begin{eqnarray}
g_{AB'CD'} = \epsilon_{AC}\:\epsilon_{ B'D'}  \ . \label{epsilon}
\end{eqnarray}

$\epsilon_{AC}$ and $\epsilon_{ B'D'}$ are the skew symmetric Levi-Civita metric spinors, both with components $\epsilon_{00} = \epsilon_{ 0'0'} = 0$,  $\epsilon_{01} = \epsilon_{ 0'1'} = 1$, $\epsilon_{11} = \epsilon_{ 1'1'} = 0$ and $\epsilon_{10} = \epsilon_{1'0'} = -1$.

\begin{eqnarray}
\zeta_{aA} = \zeta_{a}^{B}\:\epsilon_{BA} \ ,\\
\zeta_{a}^{B} = \epsilon^{BA}\:\zeta_{aA}  \ .
\end{eqnarray}

$\zeta_{a}^{B}$ are local basis spinors, the index $a$ spanning a complex two dimensional vector space ${\cal{C}}$ at each spacetime point, see reference \cite{MC}, chapter ten. In the complex conjugate to this vector space ${\cal{\overline{C}}}$ the indices are raised and lowered using the $\epsilon_{ B'D'}$ metric spinor,

\begin{eqnarray}
\eta_{aA'} = \eta_{a}^{B'}\:\epsilon_{B'A'} \ ,\\
\eta_{a}^{B'} = \epsilon^{B'A'}\:\eta_{aA'} \ .
\end{eqnarray}

Next, we write the normalization conditions,

\begin{eqnarray}
\zeta_{a}^{A}\:\epsilon_{AB}\:\zeta_{b}^{B} = \epsilon_{ab} \ , \label{normaliz}
\end{eqnarray}

and the completeness relations,

\begin{eqnarray}
\zeta_{a}^{A}\:\epsilon^{ab}\:\zeta_{b}^{B} = \epsilon^{AB} \ . \label{completit}
\end{eqnarray}

Similar results follow for the ${\cal{\overline{C}}}$ basis spinors. The local basis spinors have the following transformation law,

\begin{eqnarray}
\tilde{\zeta}_{a}^{A} = (S_{sl}^{-1})_{a}^{\:\:b}\:\zeta_{b}^{A}  \ , \label{transfbasspin}
\end{eqnarray}

where $S_{sl}$ is an $\mathrm{SL}(2,C)$ gauge transformation. Since $\epsilon^{AB}$ belongs to the tensor product ${\cal{C}} \bigotimes {\cal{C}}$, it transforms under $\mathrm{SL}(2,C)$ as,

\begin{eqnarray}
\epsilon^{'AB}  = \epsilon^{CD}\:S_{C}^{\:\:A}\:S_{D}^{\:\:B} = (det\:S)\:\epsilon^{AB}  \ , \label{transfepsilon}
\end{eqnarray}

where $\det S$ stands for the determinant of the $\mathrm{SL}(2,C)$ transformation $S_{A}^{\:\:B}$, that is $\det S = 1$. $\epsilon^{AB}$ is a metric spinor, that is why it is used for raising and lowering spinor indices. The local coordinate representation of the covariant derivative is,

\begin{eqnarray}
\nabla_{\mu}\psi^{A} = \partial_{\mu}\psi^{A} + \Gamma_{\mu\:\:\:B}^{\:\:A}\:\psi^{B} \ ,\label{covarderiv}
\end{eqnarray}

while the transformation properties of the $\Gamma_{\mu\:\:\:D}^{\:\:C}$ under $\mathrm{SL}(2,C)$ are,

\begin{eqnarray}
\tilde{\Gamma}_{\mu\:\:\:A}^{\:\:B } = (S^{-1})_{C}^{\:\:B}\:\Gamma_{\mu\:\:\:D}^{\:\:C}\:S_{A}^{\:\:D} + (S^{-1})_{C}^{\:\:B}\:\partial_{\mu}S_{A}^{\:\:C} \ .   \label{Gammatransf}
\end{eqnarray}

Next, we formulate two conditions on the metric spinors,

\begin{eqnarray}
\nabla_{\mu}\epsilon_{AB} = 0 \ .\label{ne1}
\end{eqnarray}

\begin{eqnarray}
\nabla_{\mu}\epsilon^{AB} = 0 \ .\label{ne2}
\end{eqnarray}

The two conditions (\ref{ne1}-\ref{ne2}) reduce the number of independent complex components of the spinor connection to 12 since they imply,

\begin{eqnarray}
\Gamma_{\mu\:\:\:A}^{\:\:B} = \Gamma_{\mu B}^{\:\quad A} \ .
\end{eqnarray}

Similar conditions are imposed on $\epsilon_{B'D'}$ and $\epsilon^{B'D'}$. We also demand that the operation of converting spinor to tensor indices and vice-versa commute with covariant differentiation,

\begin{eqnarray}
\nabla_{\mu}\sigma^{\nu}_{AB'} = 0 \ . \label{asf}
\end{eqnarray}

The above condition (\ref{asf}) implies that the relation between the components of the spinor connections and affine connections can be written as, see reference \cite{MC} chapter ten,

\begin{eqnarray}
\Gamma_{\mu\:\:\:A}^{\:\:C} = \frac{1}{2}\:\sigma_{\nu}^{ CB'}\:(\sigma^{\lambda}_{AB'}\:\Gamma^{\nu}_{\lambda\mu} + \partial_{\mu} \sigma^{\nu}_{AB'}) \ . \label{affineconn}
\end{eqnarray}

The gauge potentials in the $\mathrm{SL}(2,C)$ gauge theory, are the dyad components of the spinor connection,

\begin{eqnarray}
\nabla_{\mu}\zeta_{a}^{A} = \Gamma_{\mu\:\:\:B}^{\:\:A}\:\zeta_{a}^{B} \ .
\end{eqnarray}

\begin{eqnarray}
(B_{\mu})_{a}^{\:\:b} = \Gamma_{\mu\:A}^{\:\:B}\:\zeta_{a}^{A}\:\zeta^{b}_{B} \ .
\end{eqnarray}

Now, we consider the covariant derivatives of the basis vectors,

\begin{eqnarray}
\nabla_{\mu}\zeta_{a}^{A} = (B_{\mu})_{a}^{\:\:b}\:\zeta_{b}^{A} \ .
\end{eqnarray}

and the transformation of the the gauge potentials,

\begin{eqnarray}
(\tilde{B}_{\mu})_{a}^{\:\:b} = (S_{sl}^{-1})_{a}^{\:\:c}\:(B_{\mu})_{c}^{\:\:d}\:S_{sl\:d}^{\:\quad b} - (S_{sl}^{-1})_{a}^{\:\:c}\:\partial_{\mu}S_{c}^{\:\:b}  \ . \label{transfpot}
\end{eqnarray}

The field strengths or curvature tensors are given by the expression,

\begin{eqnarray}
F_{\mu\nu A}^{\:\quad B} = \partial_{\nu}\Gamma_{\mu A}^{\:\:\:\:\:B} -\partial_{\mu}\Gamma_{\nu A}^{\:\:\:\:\:B} + \Gamma_{\mu A}^{\:\:\:\:\:C}\:\Gamma_{\nu C}^{\:\:\:\:\:B} - \Gamma_{\nu  A}^{\:\:\:\:\:C}\:\Gamma_{\mu C}^{\:\:\:\:\:B}  \ , \label{fieldstrengthspin}
\end{eqnarray}

Or in dyad components,

\begin{eqnarray}
F_{\mu\nu a}^{\:\quad b} = \partial_{\nu}B_{\mu a}^{\:\:\:\:\:b} -\partial_{\mu}B_{\nu a}^{\:\:\:\:\:b} + B_{\mu a}^{\:\:\:\:\:c}\:B_{\nu c}^{\:\:\:\:\:b} - B_{\nu a}^{\:\:\:\:\:c}\:B_{\mu c}^{\:\:\:\:\:b}  \ .\label{fieldstrengthdyad}
\end{eqnarray}

Under a gauge $\mathrm{SL}(2,C)$ transformation the field strength transforms homogeneously,

\begin{eqnarray}
\tilde{F}_{\mu\nu a}^{\:\quad b} = (S_{sl}^{-1})_{a}^{\:\:c}\:F_{\mu\nu c}^{\:\quad d}\:S_{sl\:d}^{\:\quad b}  \ . \label{transffs}
\end{eqnarray}

Finally, we exhibit the relationship of the field strength and the Riemann tensor,

\begin{eqnarray}
F_{\mu\nu a}^{\:\quad b} = \frac{1}{2}\:R^{\rho}_{\:\:\lambda\mu\nu}\:\sigma_{\rho a c'}\:\sigma^{\lambda b c'} \ . \label{riemann}
\end{eqnarray}


\section{Appendix II}
\label{sec:appII}

The second appendix is introducing the object $\Sigma^{\alpha\beta}$. This object according to the matrix definitions introduced in the references is Hermitian. The use of this object in the construction of our tetrads, allows for the local $\mathrm{SL}(2,C)$ or $\mathrm{SU}(2)$ gauge transformations, to get transformed into purely local geometrical transformations. That is, local proper Lorentz transformations. The object $\sigma^{\alpha\beta}$ is defined as $\sigma^{\alpha\beta} = \sigma_{+}^{\alpha}\:\sigma_{-}^{\beta}-\sigma_{+}^{\beta}\:\sigma_{-}^{\alpha}$, \cite{MC,PR,MK,GM,MN,PS}. The object $\sigma_{\pm}^{\alpha}$ arises when building the Weyl representation for left handed and right handed spinors. According to \cite{GM}, it is defined as $\sigma_{\pm}^{\alpha} = (\bf{1},\pm\sigma^{i})$, where $\sigma^{i}$ are the Pauli matrices for $i = 1\cdots3$. Even though in reference \cite{MC} there is a different definition of these matrices, tensor expressions do not change. It must be stressed that we are suppressing either spinor or dyad indices for simplicity of notation.

It is also possible to define the object $\sigma^{\dagger\alpha\beta} = \sigma_{-}^{\alpha}\:\sigma_{+}^{\beta}-\sigma_{-}^{\beta}\:\sigma_{+}^{\alpha}$, analogously. Local $\mathrm{SU}(2)$ gauge transformations $S_{(1/2)}$ associated to local spatial rotations have already been analyzed in manuscript \cite{AYM}. Under the $(\frac{1}{2},0)$ and $(0,\frac{1}{2})$ spinor representations of the Lorentz group they transform as,

\begin{eqnarray}
S_{(1/2)}^{-1}\:\sigma^{\alpha}\:S_{(1/2)} &=& \Lambda^{\alpha}_{\:\:\:\gamma}\:\sigma^{\gamma}  \ ,\label{sigmatr1} \\
S_{(1/2)}^{-1}\:\overline{\sigma}^{\alpha}\:S_{(1/2)} &=& \Lambda^{\alpha}_{\:\:\:\gamma}\:\overline{\sigma}^{\alpha}\ .\label{sigmatr2}
\end{eqnarray}

Equations (\ref{sigmatr1}-\ref{sigmatr2}) mean that under the spinor representation of the Lorentz group, $\sigma^{\alpha} = \sigma_{+}^{\alpha}$ and $\overline{\sigma}^{\alpha} = \sigma_{-}^{\alpha}$ transform the same way as vectors. In (\ref{sigmatr1}-\ref{sigmatr2}), the matrices $S_{(1/2)}$ are local, as well as the Lorentz $\Lambda^{\alpha}_{\:\:\:\gamma}$, see reference \cite{SWB}. The $\mathrm{SU}(2)$ elements $S_{(1/2)}$ can be considered to belong to the Weyl spinor representation of the Lorentz group. Since the group $\mathrm{SU}(2)$ is homomorphic to $\mathrm{SO}(3)$, the Lorentz $\Lambda^{\alpha}_{\:\:\:\gamma}$ just represent local space rotations. For $\mathrm{SL}(2,C)$ group elements $S_{sl}$, the analogous equations to (\ref{sigmatr1}-\ref{sigmatr2}) are,

\begin{eqnarray}
S_{sl}^{-1}\:\sigma^{\alpha}\:(S_{sl}^{-1})^{\dag} &=& \Lambda^{\alpha}_{\:\:\:\gamma}\:\sigma^{\gamma}  \ , \label{sigmasl2ctr1} \\
S_{sl}^{\dag}\:\overline{\sigma}^{\alpha}\:S_{sl} &=& \Lambda^{\alpha}_{\:\:\:\gamma}\:\overline{\sigma}^{\alpha}\ .\label{sigmasl2ctr2}
\end{eqnarray}

We follow in general the notation in \cite{DHM} and since we will not write all these objects properties, we cite fundamentally the references \cite{SM,IA}. The proof to equations (\ref{sigmasl2ctr1}-\ref{sigmasl2ctr2}) is a lengthy but straightforward proof.

Nonetheless, it is relevant to understand that in order for the two vector fields $X^{\mu}$ and $Y^{\mu}$ to be real valued, we have to choose the object $\Sigma^{\alpha\beta}$ such that it is Hermitian. We will show explicitly an according to matrix definitions in \cite{MK,GM,MN,PS} the components of the following objects,

\begin{center}
$\imath \: \left(\sigma^{\alpha\beta} + \sigma^{\dagger\alpha\beta}  \right)  = \left\{ \begin{array}{ll}
				0 \:\:\:\:\: \mbox{if $\alpha = 0$ and $\beta = i$}\ ,\\
				4\:\epsilon^{ijk}\:\sigma^{k} \:\:\:\:\: \mbox{if $\alpha = i$ and $\beta = j$ \ ,}
				    \end{array}
			    \right. $
\end{center}

\begin{center}
$ \sigma^{\alpha\beta} - \sigma^{\dagger\alpha\beta}  = \left\{ \begin{array}{ll}
				-4\:\sigma^{i} \:\:\:\:\: \mbox{if $\alpha = 0$ and $\beta = i$}\ ,\\
				0 \:\:\:\:\: \mbox{if $\alpha = i$ and $\beta = j$ \ .}
				    \end{array}
			    \right. $
\end{center}

Once again we remind ourselves that we are not writing either spinor or dyad indices like $\sigma_{+}^{\mu\:AA'}$, or $\sigma_{-\:A'A}^{\mu}$, for the sake of simplicity. The reader can refer in this regard to reference \cite{MC}. We might then call $\Sigma_{ROT}^{\alpha\beta} = \imath \: \left(\sigma^{\alpha\beta} + \sigma^{\dagger\alpha\beta}  \right)$, and $\Sigma_{BOOST}^{\alpha\beta} = \sigma^{\alpha\beta} - \sigma^{\dagger\alpha\beta}$. Therefore, a possible choice for the object $\Sigma^{\alpha\beta}$ could be for instance, $\Sigma^{\alpha\beta} = \Sigma_{ROT}^{\alpha\beta}$. This a particularly suitable choice when we consider proper spatial Lorentz transformations of the tetrad vectors nested within the structure of the gauge vectors $X^{\mu}$ and $Y^{\mu}$.


The gauge covariant derivatives are telling us the following \cite{MK,MN,PS}. The gauge covariant derivative ${\cal{D}}_{\mu} \psi = \partial_{\mu}\psi_{AM} - \Gamma_{\mu M}^{\:\:\:\:\:\:Q}\:\psi_{AQ} - \imath \:g\:A_{\mu A}^{\:\:\:\:\:\:Q}\:\psi_{QM} + \imath\:\frac{g^{'}}{2}\:B^{Y}_{\mu}\:\psi_{AM} = \partial_{\mu}\:\psi_{AM} + \frac{1}{2}\:\left( \sigma^{\beta\gamma} \right)_{M}^{\:\:\:\:Q}\:V_{\beta}^{\nu}\: V_{\gamma\nu;\mu}\:\psi_{AQ} - \imath \:g\:A_{\mu A}^{\:\:\:\:\:\:Q}\:\psi_{QM} + \imath\:\frac{g^{'}}{2}\:B^{Y}_{\mu}\:\psi_{AM}$, commutes with gauge tranformations, in this case local $\mathrm{U}(1)$, $\mathrm{SU}(2)$ and $\mathrm{SL}(2,C)$ gauge transformations \cite{MC,PR,MK,GM,MN,PS,SWB,DHM,SM,IA,FQ,MS,JP,RU,YM,WE}. That means, it commutes with LB1 and LB2 transformations, or in other words, the generating vectors of blades one and two at each spacetime point are not distinguished by the derivative when ``rotated'' inside the blades they generate, because the derivative commutes with the local ``rotation'' itself. That is the geometrical meaning of the gauge covariant derivative when commuting with either $\mathrm{U}(1)$, $\mathrm{SU}(2)$ or $\mathrm{SL}(2,C)$ local gauge transformations. We must notice that $\Gamma_{\mu M}^{\:\:\:\:\:\:Q}$ are the spinor components of the $\mathrm{SL}(2,C)$ potential. It can be written as in expression (\ref{affineconn}) or in the alternative form $\frac{1}{2}\:\left( \sigma^{\beta\gamma} \right)_{M}^{\:\:\:\:Q}\:V_{\beta}^{\nu}\: V_{\gamma\nu;\mu}$ see reference \cite{RU,WE} (the signature in \cite{WE} is -+++). The form we choose depends on the variables we are using. We can also think the two component $\mathrm{SU}(2)$ basis spinor fields $\eta_{p}^{P}$, as right handed. We might analogously implement a formalism for left handed spinor fields \cite{GR2,MC}. We must also stress that the spinors $\psi_{AM}$ are not necessarily the tensor product of a $\mathrm{SU}(2)$ spinor $\eta_{A}$ and a $\mathrm{SL}(2,C)$ spinor $\zeta_{M}$, that is $\psi_{AM} \neq \eta_{A}\:\zeta_{M}$ in general. $\mathrm{SL}(2,C) \times \mathrm{SU}(2)$ spinors are written in terms of the tensor product basis $\eta_{A}\:\zeta_{M}$ see chapter ten in \cite{MC}.


\section{Appendix III}
\label{sec:appIII}

Let us define an extremal field for non-Abelian theories as,

\begin{equation}
\zeta_{\mu\nu} = \cos\beta \:\: f_{\mu\nu}-
\sin\beta \:\: \ast f_{\mu\nu} \ ,\label{exsu2}
\end{equation}

In order to define the new complexion $\beta$ we will impose the new $\mathrm{SU}(2)$ local invariant condition,

\begin{eqnarray}
\mathrm{Tr}[\zeta_{\mu\nu}\:\ast \zeta^{\mu\nu}]=\zeta^{k}_{\mu\nu}\:\ast \zeta^{k\mu\nu} &=& 0\ ,\label{ccsu2}
\end{eqnarray}

where the summation convention has been applied on the internal index $k$. The complexion condition (\ref{ccsu2}) is not an additional condition for the field strength. We are just using a generalized duality transformation, and defining through it, this new local scalar complexion $\beta$. We simply introduced a possible generalization of the definition for the Abelian complexion, found through a duality transformation as well. Then, the local $\mathrm{SU}(2)$ invariant complexion $\beta$ turns out to be,

\begin{eqnarray}
\tan(2\beta) = - f^{k}_{\mu\nu}\:\ast f^{k\mu\nu} / f^{p}_{\lambda\rho}\:f^{p\lambda\rho}\ ,\label{compksu2}
\end{eqnarray}

where again the summation convention was applied on both $k$ and $p$. Now we would like to consider gauge covariant derivatives since they will become useful afterwards. For instance, the gauge covariant derivatives of the three extremal field internal components,

\begin{eqnarray}
\zeta_{k\mu\nu\mid\rho} = \zeta_{k\mu\nu\, ; \, \rho} + g \: \epsilon_{klp}\: A_{l\rho}\:\zeta_{p\mu\nu}\ .\label{gcd}
\end{eqnarray}

where $\epsilon_{klp}$ is the completely skew-symmetric tensor in three dimensions with $\epsilon_{123} = 1$, and where $g$ is the coupling constant. The symbol ``;'' stands for the usual covariant derivative associated with the metric tensor $g_{\mu\nu}$. We will call our geometrized electromagnetic potential $A^{\alpha}$, where $f_{\mu\nu}=A_{\nu ;\mu} - A_{\mu ;\nu}$ is the geometrized electromagnetic field $f_{\mu\nu}= (G^{1/2} / c^2) \: F_{\mu\nu}$. Analogously, $f^{k}_{\mu\nu}$ are the geometrized Yang-Mills field components, $f^{k}_{\mu\nu}= (G^{1/2} / c^2) \: F^{k}_{\mu\nu}$. We will consider for instance the Einstein-Maxwell-Yang-Mills vacuum field equations,

\begin{eqnarray}
R_{\mu\nu} &=& T^{(ym)}_{\mu\nu} + T^{(em)}_{\mu\nu}\ ,\label{eyme2}\\
f^{\mu\nu}_{\:\:\:\:\:;\nu} &=& 0 \ ,\label{EM12}\\
\ast f^{\mu\nu}_{\:\:\:\:\:;\nu} &=& 0 \ ,\label{EM22}\\
f^{k\mu\nu}_{\:\:\quad\mid \nu} &=& 0 \ ,\label{ymvfe12}\\
\ast f^{k\mu\nu}_{\:\:\quad\mid \nu} &=& 0 \ . \label{ymvfe22}
\end{eqnarray}

The field equations (\ref{EM12}-\ref{EM22}) provide a hint about the existence of two electromagnetic field potentials \cite{CF}, as said in the first paper ``Tetrads in geometrodynamics'', not independent from each other, but due to the symmetry of the equations, available for our construction. $A^{\mu}$ and $\ast A^{\mu}$ are the two electromagnetic potentials. $\ast A^{\mu}$ is therefore a name, we are not using the Hodge map at all in this case, meaning that $\ast A_{\nu ;\mu} = (\ast A_{\nu})_{;\mu}$. These two potentials are not independent from each other, nonetheless they exist and are available for our construction. Similar for the two Yang-Mills Non-Abelian  equations (\ref{ymvfe12}-\ref{ymvfe22}). The Non-Abelian potential $A^{k\mu}$ is available for our construction as well \cite{MC,YM,RU}. We briefly can remind ourselves from reference \cite{gaugeinvmeth} that we can introduce a generalized duality transformation for Yang-Mills non-Abelian fields. For instance we might choose,

\begin{equation}
\varepsilon_{\mu\nu} =  \mathrm{Tr}[\vec{m}\: \cdot \: f_{\mu\nu} - \vec{l}\: \cdot \: \ast f_{\mu\nu}] \ ,\label{gendual}
\end{equation}

where the field strength $f_{\mu\nu} = f^{a}_{\mu\nu}\:\sigma^{a}$, and $\vec{m} = m^{a}\:\sigma^{a}$ and $\vec{l} = l^{a}\:\sigma^{a}$ are vectors in isospace. The $\cdot$ means again product in isospace. Once more we stress that $\sigma^{a}$ are the Pauli matrices and the summation convention is applied on the internal index $a$. The vector components are defined as,

\begin{eqnarray}
\lefteqn{ \vec{m} = (\cos\alpha_{1},\cos\alpha_{2},\cos\alpha_{3})\ , } \label{ISO1} \\
&&\vec{l} = (\cos\beta_{1},\cos\beta_{2},\cos\beta_{3}) \ , \label{ISO2}
\end{eqnarray}

where all the six isoangles are local scalars that satisfy,

\begin{eqnarray}
\lefteqn{ \Sigma_{a=1}^{3} \cos^{2}\alpha_{a} = 1 \ ,} \label{ISOSUM1} \\
&&\Sigma_{a=1}^{3} \cos^{2}\beta_{a} = 1 \ . \label{ISOSUM2}
\end{eqnarray}

In isospace $\vec{m} = m^{a}\:\sigma^{a}$ transforms under a local $\mathrm{SU}(2)$ gauge transformation $S$, as $S^{-1}\:\vec{m}\:S$, see chapter III in \cite{CBDW} and also reference \cite{GRSYMM}, and similar for $\vec{l} = l^{a}\:\sigma^{a}$. The tensor $f_{\mu\nu} = f^{a}_{\mu\nu}\:\sigma^{a}$ transforms as
$f_{\mu\nu} \rightarrow S^{-1}\:f_{\mu\nu}\:S$. Therefore $\varepsilon_{\mu\nu}$ is manifestly gauge invariant. We can see from (\ref{ISO1}-\ref{ISO2}) and (\ref{ISOSUM1}-\ref{ISOSUM2}) that only four of the six angles in isospace are independent. Once our choice is made, then the duality rotation we perform next in order to obtain the new extremal field is given by,

\begin{eqnarray}
\epsilon_{\mu\nu} = \cos\vartheta \: \varepsilon_{\mu\nu} - \sin\vartheta \:
\ast \varepsilon_{\mu\nu}\ .\label{extremalR}
\end{eqnarray}

As always we choose this complexion $\vartheta$ to be defined by the condition,

\begin{eqnarray}
\epsilon_{\mu\nu}\:\ast \epsilon^{\mu\nu} &=& 0\ ,\label{rc}
\end{eqnarray}

which implies that,

\begin{eqnarray}
\tan(2\vartheta) = - \varepsilon_{\mu\nu}\:\ast \varepsilon^{\mu\nu} / \varepsilon_{\lambda\rho}\:\varepsilon^{\lambda\rho}\ .\label{compr}
\end{eqnarray}

We proceed then to pick up the only $\mathrm{SU}(2)$ tetrad involved in our new proof. Following the same procedure depicted in manuscript \cite{AYM} section ``Extremal field in $\mathrm{SU}(2)$ geometrodynamics'' we introduce using the same notation, the tetrad,

\begin{eqnarray}
S_{(1)}^{\mu} &=& \epsilon^{\mu\lambda}\:\epsilon_{\rho\lambda}\:X^{\rho}\ ,
\label{S1app}\\
S_{(2)}^{\mu} &=& \sqrt{-Q_{ym}/2} \: \epsilon^{\mu\lambda} \: X_{\lambda}\ ,
\label{S2app}\\
S_{(3)}^{\mu} &=& \sqrt{-Q_{ym}/2} \: \ast \epsilon^{\mu\lambda} \: Y_{\lambda}\ ,
\label{S3app}\\
S_{(4)}^{\mu} &=& \ast \epsilon^{\mu\lambda}\: \ast\epsilon_{\rho\lambda}
\:Y^{\rho}\ ,\label{S4app}
\end{eqnarray}

where $Q_{ym} = \epsilon_{\mu\nu}\:\epsilon^{\mu\nu}$ is assumed not to be zero and $\epsilon_{\mu\nu}$ is a new kind of local $\mathrm{SU}(2)$ gauge invariant extremal tensor already introduced above and also in paper \cite{AYM} obeying the local property $\epsilon_{\alpha\mu}\:\ast \epsilon^{\mu\nu} = 0$. The expression  $\ast \epsilon_{\mu\nu}={1 \over 2}\:\epsilon_{\mu\nu\sigma\tau}\:\epsilon^{\sigma\tau}$ is the dual tensor of an antisymmetric tensor $\epsilon_{\mu\nu}$. When we use the general identity (\ref{2ig}) which is valid for every pair of antisymmetric tensors in a four-dimensional Lorentzian spacetime \cite{MW}, and apply it to the case
$A_{\mu\alpha} = \epsilon_{\mu\alpha}$ and $B^{\nu\alpha} = \ast \epsilon^{\nu\alpha}$, it yields an equivalent condition to (\ref{rc}),

\begin{eqnarray}
\epsilon_{\alpha\mu}\:\ast \epsilon^{\mu\nu} &=& 0\ .\label{rceqna}
\end{eqnarray}

\section{Appendix IV: Kernel of the mapping}
\label{sec:appIV}

\subsection{Kernel on plane one}
\label{kernelp1}

Since we are focused on the electromagnetic local gauge transformations of the tetrad vectors,

\begin{eqnarray}
V_{(1)}^{\alpha} &=& \xi^{\alpha\lambda}\:\xi_{\rho\lambda}\:A^{\rho}\ ,
\label{V1}\\
V_{(2)}^{\alpha} &=& \sqrt{-Q/2} \: \xi^{\alpha\lambda} \: A_{\lambda}\ ,
\label{V2}\\
V_{(3)}^{\alpha} &=& \sqrt{-Q/2} \: \ast \xi^{\alpha\lambda}
\: \ast A_{\lambda}\ ,\label{V3}\\
V_{(4)}^{\alpha} &=& \ast \xi^{\alpha\lambda}\: \ast \xi_{\rho\lambda}
\:\ast A^{\rho}\ ,\label{V4}
\end{eqnarray}

we set out now to study the Kernel of this mapping. The Kernel of the transformations $X_{\alpha} = A_{\alpha} \mapsto  A_{\alpha} + \Lambda_{,\alpha}$ on the local plane one is determined  by equations (58-59) in \cite{A},

\begin{eqnarray}
{\tilde{V}_{(1)}^{\alpha}
\over \sqrt{-\tilde{V}_{(1)}^{\beta}\:\tilde{V}_{(1)\beta}}}&=&
{(1+C) \over \sqrt{(1+C)^2-D^2}}
\:{V_{(1)}^{\alpha} \over \sqrt{-V_{(1)}^{\beta}\:V_{(1)\beta}}}+
{D \over \sqrt{(1+C)^2-D^2}}
\:{V_{(2)}^{\alpha} \over \sqrt{V_{(2)}^{\beta}\:V_{(2)\beta}}}\ ,\label{TN1N}\\
{\tilde{V}_{(2)}^{\alpha}
\over \sqrt{\tilde{V}_{(2)}^{\beta}\:\tilde{V}_{(2)\beta}}}&=&
{D \over \sqrt{(1+C)^2-D^2}}
\:{V_{(1)}^{\alpha} \over \sqrt{-V_{(1)}^{\beta}\:V_{(1)\beta}}} +
{(1+C) \over \sqrt{(1+C)^2-D^2}}
\:{V_{(2)}^{\alpha} \over \sqrt{V_{(2)}^{\beta}\:V_{(2)\beta}}}\ .
\label{TN2N}
\end{eqnarray}

In this manuscript is given by the expressions,

\begin{eqnarray}
C + 1 &=&(-Q/2)\:V_{(1)\sigma}\:\Lambda^{\sigma} / (\:V_{(2)\beta}\:
V_{(2)}^{\beta}\:) + 1 > 0 \ ,\label{COEFFCKER}\\
D &=&(-Q/2)\:V_{(2)\sigma}\:\Lambda^{\sigma} / (\:V_{(1)\beta}\:
V_{(1)}^{\beta}\:) = 0 \ .\label{COEFFDKER}
\end{eqnarray}

Equations (\ref{COEFFCKER}-\ref{COEFFDKER}) reduce to $C+1 > 0$ and $V_{(2)\sigma}\:\Lambda^{\sigma} = 0$. This is the Kernel of the mapping between the local group of electromagnetic gauge transformations and the local group of tetrad transformations on the local blade or plane one. The equation $V_{(2)\sigma}\:\Lambda^{\sigma} = 0$ establishes that the Kernel on plane one is given by the local gauge transformations $\Lambda^{\sigma}$ that have no projection on the $V_{(2)\sigma}$ tetrad vector. That is,

\begin{eqnarray}
V_{(2)\sigma}\:\Lambda^{\sigma} = 0 \: \Leftrightarrow \: A^{\tau}\:\xi_{\tau\sigma}\:\Lambda^{\sigma} = 0 \ . \label{ORTHOKER}
\end{eqnarray}

This equation (\ref{ORTHOKER}) implies that both $\Lambda^{\sigma}$ and $A^{\tau}$ have zero projection on the tetrad vector $V_{(2)}^{\sigma}$ because $V_{(2)\sigma}\:A^{\sigma} = A^{\sigma}\:\xi_{\sigma\tau}\:A^{\tau} = 0$. Therefore, on plane one $\Lambda^{\sigma}$ and $A^{\tau}$ have local proportionality between their projections on the local tetrad vector $V_{(1)}^{\sigma}$ or its normalized version $U^{\sigma}$. We remind ourselves that the normalized version of vectors (\ref{V1}-\ref{V4}) is given by the tetrad vectors (\ref{U}-\ref{W}), where we have assumed that $- V_{(1)}^{\alpha}\:V_{(1)\alpha}= V_{(2)}^{\alpha}\:V_{(2)\alpha} > 0$ for simplicity. Then,

\begin{eqnarray}
C\:U^{\tau}\:A_{\tau} = U^{\tau}\:\Lambda_{\tau} \ , \label{POPORBLADEONE}
\end{eqnarray}

just because both vectors only have non-zero projection on the local plane one vector $V_{(1)}^{\tau}$ or the normalized $U^{\tau}$. It is a matter of simple algebra to prove that C is just the local scalar presented previously as in equation (\ref{COEFFCKER}). Therefore we could in general in all the spacetime write the following equality,

\begin{eqnarray}
C\:\:A^{\mu} = \Lambda^{\mu} + R_{1}\:V_{(3)}^{\mu} + S_{1}\:V_{(4)}^{\mu} \ ,\label{ASUM}
\end{eqnarray}

where $R_{1}$ and $S_{1}$ are just local scalars and ($V_{(3)}^{\mu} , V_{(4)}^{\mu}$) span or generate the local plane two, orthogonal to plane one.
We can rewrite equation (\ref{ASUM}) as,

\begin{eqnarray}
\Lambda^{\mu} =  C\:A^{\mu} - R_{1}\:V_{(3)}^{\mu} - S_{1}\:V_{(4)}^{\mu} \ .\label{ASUM2}
\end{eqnarray}

Therefore we have found in principle that the Kernel is associated to three scalars C,$R_{1}$ and $S_{1}$. In equation (\ref{ASUM2}) the potential vector $A^{\mu}$, and the tetrad vectors $V_{(3)}^{\mu}$ and $V_{(4)}^{\mu}$ are supposed to be known. This is all just appearance of a Kernel because the local gradient of a scalar $\Lambda^{\mu}$ obeys six equations or conditions of integrability plus the condition (\ref{COEFFCKER}) $1+C>0$. These are seven conditions for three local scalars. In reality we should be considering the three local scalars as functions of four coordinates each and therefore there will be three gradients or twelve derivatives embedded in the integrability conditions. So, in principle the Kernel will not be trivial while at the same time we can say more about this Kernel in order to simplify its structure. When we replace equation (\ref{ASUM}) back in vectors (\ref{V1}-\ref{V2}) we obtain,

\begin{eqnarray}
V_{(1)}^{\alpha} &=& \xi^{\alpha\lambda}\:\xi_{\rho\lambda}\:[A^{\rho}] = \xi^{\alpha\lambda}\:\xi_{\rho\lambda}\:[\frac{1}{C}\:\Lambda^{\rho} - \frac{R_{1}}{C}\:V_{(3)}^{\rho} - \frac{S_{1}}{C}\:V_{(4)}^{\rho}]\ ,
\label{V1PUREGAUGE}\\
V_{(2)}^{\alpha} &=& \sqrt{-Q/2} \: \xi^{\alpha\lambda} \: [A_{\lambda}] = \sqrt{-Q/2} \: \xi^{\alpha\lambda} \: [\frac{1}{C}\:\Lambda_{\lambda} - \frac{R_{1}}{C}\:V_{(3)\lambda} - \frac{S_{1}}{C}\:V_{(4)\lambda}]  \ .
\label{V2PUREGAUGE}
\end{eqnarray}

since we can use repeatedly equation (\ref{rceq}) we find,

\begin{eqnarray}
V_{(1)}^{\alpha} &=& \xi^{\alpha\lambda}\:\xi_{\rho\lambda}\:[\frac{1}{C}\:\Lambda^{\rho}] = \frac{1}{C}\:\xi^{\alpha\lambda}\:\xi_{\rho\lambda}\:[\Lambda^{\rho}]\ ,
\label{V1PUREGAUGE2}\\
V_{(2)}^{\alpha} &=& \sqrt{-Q/2} \: \xi^{\alpha\lambda} \: [\frac{1}{C}\:\Lambda_{\lambda}] = \frac{1}{C}\:\sqrt{-Q/2} \: \xi^{\alpha\lambda} \: [\Lambda_{\lambda}] \ .
\label{V2PUREGAUGE2}
\end{eqnarray}

Then, we find for the normalized version of (\ref{V1PUREGAUGE2}-\ref{V2PUREGAUGE2}),

\begin{eqnarray}
U^{\alpha} &=& {|C| \over C}\:\xi^{\alpha\lambda}\:\xi_{\rho\lambda}\:\Lambda^{\rho} \:
/ \: (\: \sqrt{-Q/2} \: \sqrt{\Lambda_{\mu} \ \xi^{\mu\sigma} \
\xi_{\nu\sigma} \ \Lambda^{\nu}}\:) \ ,\label{UPUREGAUGE}\\
V^{\alpha} &=& {|C| \over C}\:\xi^{\alpha\lambda}\:\Lambda_{\lambda} \:
/ \: (\:\sqrt{\Lambda_{\mu} \ \xi^{\mu\sigma} \
\xi_{\nu\sigma} \ \Lambda^{\nu}}\:) \ , \label{VPUREGAUGE}
\end{eqnarray}

assuming that $\Lambda_{\mu} \: \xi^{\mu\sigma} \: \xi_{\nu\sigma} \: \Lambda^{\nu} > 0$ for simplicity of analysis. It is as if we chose a local gauge vector which is pure gauge $X^{\lambda}=\pm\:\Lambda^{\lambda}$ even though we know that the whole expression for $A^{\rho}$ might include a component on plane two, orthogonal to plane one as in equation (\ref{ASUM}). This orthogonality is manifested by the equation (\ref{rceq}) since we already know that $\xi_{\mu\nu}\:V_{(3)}^{\mu}=0$ and $\xi_{\mu\nu}\:V_{(4)}^{\mu}=0$, specially in equations (\ref{V1PUREGAUGE}-\ref{V2PUREGAUGE}) after replacing by equation (\ref{ASUM}). The second case would be a gradient $\Lambda^{\mu}=R_{1}\:V_{(3)}^{\mu} + S_{1}\:V_{(4)}^{\mu}$ that is just a vector in the local plane two without component in the local plane one. In this case we would have genuine gauge transformations in the Kernel. Let us call $\mathrm{PGB2}=\{\Lambda / \Lambda^{\mu} \in \mbox{local Plane 2} \}$ the set of pure gauge in blade two. This would be the Kernel for the mapping between $\mathrm{U}(1)$ and LB1. We anticipate that we will have an isomorphism between the group of local electromagnetic gauge transformations minus the local group $\mathrm{PGB2}$ and the local group LB1 of tetrad transformation in plane one. Because the group $\mathrm{PGB2}$ is of measure zero. It is just gradients of scalars in a local plane in a four-dimensional spacetime. If we replace equation (\ref{ASUM2}) in the local gauge transformation of vectors (\ref{U}-\ref{V}) we find the following transformed vectors

\begin{eqnarray}
\widetilde{U}^{\alpha} &=& \frac{(1+C)}{|1+C|}\:\xi^{\alpha\lambda}\:\xi_{\rho\lambda}\:A^{\rho} \:
/ \: (\: \sqrt{-Q/2} \: \sqrt{A_{\mu} \ \xi^{\mu\sigma} \
\xi_{\nu\sigma} \ A^{\nu}}\:) = \frac{(1+C)}{|1+C|}\:U^{\alpha} \ ,\label{UTR}\\
\widetilde{V}^{\alpha} &=& \frac{(1+C)}{|1+C|}\:\xi^{\alpha\lambda}\:A_{\lambda} \:
/ \: (\:\sqrt{A_{\mu} \ \xi^{\mu\sigma} \
\xi_{\nu\sigma} \ A^{\nu}}\:) = \frac{(1+C)}{|1+C|}\:V^{\alpha}  \ . \label{VTR}
\end{eqnarray}

Therefore, from equations (\ref{UTR}-\ref{VTR}) we infer that there are two possibilities, either $1+C>0$ or $1+C<0$. If $1+C<0$ we would obtain a local tetrad transformation represented by the full inversion or minus the identity two by two which is a valid case outside the Kernel. If the case is that $1+C>0$ then the local tetrad transformation on plane one would truly be in the Kernel fulfilling equations (\ref{COEFFCKER}-\ref{COEFFDKER}). This is a case where the projection of equation (\ref{ASUM2}) on plane one inside the vectors (\ref{V1}-\ref{V2}) would result in $\Lambda^{\mu}=C\:A^{\mu}$ and we know that the gradient of $\Lambda$ in equation (\ref{ASUM2}) satisfies six integrability conditions. It is as if we chose the local gauge transformation to be $\Lambda^{\mu}=C\:A^{\mu}$. Immediately, we notice from equations (\ref{V1PUREGAUGE2}-\ref{V2PUREGAUGE2}) and (\ref{UPUREGAUGE}-\ref{VPUREGAUGE}) that this kind of local gauge transformations are associated to projections of vector potentials that are related to the gradient of the local gauge transformation scalar through a local scalar factor which happens to be $\frac{1}{C}$ as in equation (\ref{ASUM}). These would be gauge transformations associated to a potential that is pure gauge by itself and we are not considering these cases.

\subsection{Kernel on plane two}
\label{kernelp2}

There is a whole similar analysis on the local plane two for the mapping $Y_{\alpha} = \ast A_{\alpha} \mapsto  \ast A_{\alpha} + \ast \Lambda_{,\alpha}$ between the local group of electromagnetic gauge transformations and the local group of tetrad transformations on the local plane two corresponding to equations (91-92) in reference \cite{A}, that is,

\begin{eqnarray}
{\tilde{V}_{(3)}^{\alpha}
\over \sqrt{\tilde{V}_{(3)}^{\beta}\:\tilde{V}_{(3)\beta}}}&=&
{(1+N) \over \sqrt{(1+N)^2+M^2}}
\:{V_{(3)}^{\alpha} \over \sqrt{V_{(3)}^{\beta}\:V_{(3)\beta}}} -
{M \over \sqrt{(1+N)^2+M^2}}
\:{V_{(4)}^{\alpha} \over \sqrt{V_{(4)}^{\beta}\:V_{(4)\beta}}}\ ,\label{TN3}\\
{\tilde{V}_{(4)}^{\alpha}
\over \sqrt{\tilde{V}_{(4)}^{\beta}\:\tilde{V}_{(4)\beta}}}&=&
{M \over \sqrt{(1+N)^2+M^2}}
\:{V_{(3)}^{\alpha} \over \sqrt{V_{(3)}^{\beta}\:V_{(3)\beta}}} +
{(1+N) \over \sqrt{(1+N)^2+M^2}}
\:{V_{(4)}^{\alpha} \over \sqrt{V_{(4)}^{\beta}\:V_{(4)\beta}}}\ .
\label{TN4}
\end{eqnarray}

As long as $[(1+N)^2+M^2]>0$ the transformations (\ref{TN3}-\ref{TN4}) are telling us that an electromagnetic gauge transformation on the vector field $\ast A^{\alpha}$ that leaves invariant the dual electromagnetic field $\ast f_{\mu\nu}=(\ast A_{\nu})_{;\mu} - (\ast A_{\mu})_{;\nu}$, generates a spatial rotation on the normalized tetrad vector fields $\left({V_{(3)}^{\alpha} \over \sqrt{V_{(3)}^{\beta}\:V_{(3)\beta}}},
{V_{(4)}^{\alpha} \over \sqrt{V_{(4)}^{\beta}\:V_{(4)\beta}}}\right)$. The Kernel conditions on plane two are given by,

\begin{eqnarray}
N + 1 &=&(-Q/2)\:V_{(4)\sigma}\:\ast \Lambda^{\sigma} / (\:V_{(3)\beta}\:
V_{(3)}^{\beta}\:) + 1 > 0 \ ,\label{COEFFNKER}\\
M &=&(-Q/2)\:V_{(3)\sigma}\:\ast \Lambda^{\sigma} / (\:V_{(4)\beta}\:
V_{(4)}^{\beta}\:) = 0 \ .\label{COEFFMKER}
\end{eqnarray}

Equations (\ref{COEFFNKER}-\ref{COEFFMKER}) reduce to $N+1 > 0$ and $V_{(3)\sigma}\:\ast \Lambda^{\sigma} = 0$. This is the Kernel of the mapping between the local group of electromagnetic gauge transformations and the local group of tetrad transformations on the local blade or plane two. The equation $V_{(3)\sigma}\:\ast \Lambda^{\sigma} = 0$ establishes that the Kernel on plane two is given by the local gauge transformations $\ast \Lambda^{\sigma}$ that have no projection on the $V_{(3)\sigma}$ tetrad vector. That is,

\begin{eqnarray}
V_{(3)\sigma}\:\ast \Lambda^{\sigma} = 0 \: \Leftrightarrow \: \ast A^{\tau}\:\ast \xi_{\tau\sigma}\:\ast \Lambda^{\sigma} = 0 \ . \label{ORTHOKER2}
\end{eqnarray}

This equation (\ref{ORTHOKER2}) implies that both $\ast \Lambda^{\sigma}$ and $\ast A^{\tau}$ have zero projection on the tetrad vector $V_{(3)}^{\sigma}$ because $V_{(3)\sigma}\:\ast A^{\sigma} = \ast A^{\sigma}\:\ast \xi_{\sigma\tau}\:\ast A^{\tau} = 0$. Therefore, on plane two $\ast \Lambda^{\sigma}$ and $\ast A^{\tau}$ have local proportionality between their projections on the local tetrad vector $V_{(4)}^{\sigma}$ or its normalized version $W^{\sigma}$. We remind ourselves that the normalized version of vectors (\ref{V1}-\ref{V4}) is given by (\ref{U}-\ref{W}), where we have assumed that $V_{(3)}^{\alpha}\:V_{(3)\alpha}= V_{(4)}^{\alpha}\:V_{(4)\alpha} > 0$ for simplicity. Then,

\begin{eqnarray}
N\:\:W^{\tau}\:\ast A_{\tau} = W^{\tau}\:\ast \Lambda_{\tau} \ , \label{POPORBLADETWO}
\end{eqnarray}

just because both vectors only have non-zero projection on the local plane two vector $V_{(4)}^{\tau}$ or the normalized $W^{\tau}$. It is a matter of simple algebra to prove that N in equation (\ref{POPORBLADETWO}) is just the local scalar presented previously as in equation (\ref{COEFFNKER}). Therefore we could in general in all the spacetime write the following equality,

\begin{eqnarray}
N\:\:\ast A^{\mu} = \ast \Lambda^{\mu} + R_{2}\:V_{(1)}^{\mu} + S_{2}\:V_{(2)}^{\mu} \ ,\label{ASUM3}
\end{eqnarray}

where $R_{2}$ and $S_{2}$ are just local scalars and ($V_{(1)}^{\mu} , V_{(2)}^{\mu}$) span or generate the local plane one, orthogonal to plane two. We can rewrite equation (\ref{ASUM3}) as,

\begin{eqnarray}
\ast \Lambda^{\mu} =  N\:\ast A^{\mu} - R_{2}\:V_{(1)}^{\mu} - S_{2}\:V_{(2)}^{\mu} \ .\label{ASUM4}
\end{eqnarray}

Therefore we have found in principle that the Kernel is associated to three scalars N,$R_{2}$ and $S_{2}$. In equation (\ref{ASUM4}) the potential vector $\ast A^{\mu}$, and the tetrad vectors $V_{(1)}^{\mu}$ and $V_{(2)}^{\mu}$ are supposed to be known. This is all just appearance of a Kernel because the local gradient of a scalar $\ast \Lambda^{\mu}$ obeys six equations or conditions of integrability plus the condition (\ref{COEFFNKER}) $1+N>0$. These are seven conditions for three local scalars. In reality we should be considering the three local scalars as functions of four coordinates each and therefore there will be three gradients or twelve derivatives embedded in the integrability conditions. So, in principle the Kernel will not be trivial while at the same time we can say more about this Kernel in order to simplify its structure. When we replace equation (\ref{ASUM3}) back in vectors (\ref{V3}-\ref{V4}) we obtain,

\begin{eqnarray}
V_{(3)}^{\alpha} &=& \sqrt{-Q/2} \: \ast \xi^{\alpha\lambda} \: [\ast A_{\lambda}] \nonumber\\&=& \sqrt{-Q/2} \: \ast \xi^{\alpha\lambda} \: [\frac{1}{N}\:\ast \Lambda_{\lambda} - \frac{R_{2}}{N}\:V_{(1)\lambda} - \frac{S_{2}}{N}\:V_{(2)\lambda}] \ , \label{V3PUREGAUGE}\\
V_{(4)}^{\alpha} &=& \ast \xi^{\alpha\lambda}\:\ast \xi_{\rho\lambda}\:[\ast A^{\rho}] = \ast \xi^{\alpha\lambda}\:\ast \xi_{\rho\lambda}\:[\frac{1}{N}\:\ast \Lambda^{\rho} - \frac{R_{2}}{N}\:V_{(1)}^{\rho} - \frac{S_{2}}{N}\:V_{(2)}^{\rho}] \ .
\label{V4PUREGAUGE}
\end{eqnarray}

since we can use repeatedly equation (\ref{rceq}) we find,

\begin{eqnarray}
V_{(3)}^{\alpha} &=& \sqrt{-Q/2} \: \ast \xi^{\alpha\lambda} \: [\frac{1}{N}\:\ast \Lambda_{\lambda}] = \frac{1}{N}\:\sqrt{-Q/2} \: \ast \xi^{\alpha\lambda} \: [\ast \Lambda_{\lambda}] \ ,\label{V3PUREGAUGE2} \\
V_{(4)}^{\alpha} &=& \ast \xi^{\alpha\lambda}\:\ast \xi_{\rho\lambda}\:[\frac{1}{N}\:\ast \Lambda^{\rho}] = \frac{1}{N}\:\ast \xi^{\alpha\lambda}\:\ast \xi_{\rho\lambda}\:[\ast \Lambda^{\rho}] \ .
\label{V4PUREGAUGE2}
\end{eqnarray}

Then, we find for the normalized version of (\ref{V3PUREGAUGE2}-\ref{V4PUREGAUGE2}),

\begin{eqnarray}
Z^{\alpha} &=& {|N| \over N}\:\ast \xi^{\alpha\lambda}\:\ast \Lambda_{\lambda} \:
/ \: (\:\sqrt{\Lambda_{\mu} \ \ast \xi^{\mu\sigma} \
\ast \xi_{\nu\sigma} \ \ast \Lambda^{\nu}}\:)\ , \label{ZPUREGAUGE}\\
W^{\alpha} &=& {|N| \over N}\:\ast \xi^{\alpha\lambda}\:\ast \xi_{\rho\lambda}\:\ast \Lambda^{\rho} \:
/ \: (\: \sqrt{-Q/2} \: \sqrt{\ast \Lambda_{\mu} \ \ast \xi^{\mu\sigma} \
\ast \xi_{\nu\sigma} \ \ast \Lambda^{\nu}}\:) \ ,\label{WPUREGAUGE}
\end{eqnarray}

assuming that $\ast \Lambda_{\mu} \: \ast \xi^{\mu\sigma} \: \ast \xi_{\nu\sigma} \: \ast \Lambda^{\nu} > 0$ for simplicity of analysis. It is as if we chose a local gauge vector which is pure gauge $Y^{\lambda}=\pm\:\ast \Lambda^{\lambda}$ even though we know that the whole expression for $A^{\rho}$ might include a component on plane one, orthogonal to plane two as in equation (\ref{ASUM3}). This orthogonality is manifested by the equation (\ref{rceq}) since we already know that $\ast \xi_{\mu\nu}\:V_{(1)}^{\mu}=0$ and $\ast \xi_{\mu\nu}\:V_{(2)}^{\mu}=0$, specially in equations (\ref{V3PUREGAUGE}-\ref{V4PUREGAUGE}) after replacing by equation (\ref{ASUM3}). The second case would be a gradient $\Lambda^{\mu}=R_{2}\:V_{(1)}^{\mu} + S_{2}\:V_{(2)}^{\mu}$ that is just a vector in the local plane one without component in the local plane two. In this case we would have genuine gauge transformations in the Kernel. Let us call $\mathrm{PGB1}=\{\Lambda / \Lambda^{\mu} \in \mbox{local Plane 1} \}$ the set of pure gauge in blade one. This would be the Kernel for the mapping between $\mathrm{U}(1)$ and LB2. We anticipate that we will have an isomorphism between the group of local electromagnetic gauge transformations minus the local group $\mathrm{PGB1}$ and the local group LB2 of tetrad transformation in plane two. Because the group $\mathrm{PGB1}$ is of measure zero. It is just gradients of scalars in a local plane in a four-dimensional spacetime. If we replace equation (\ref{ASUM3}) in the local gauge transformation of vectors (\ref{Z}-\ref{W}) we find the following transformed vectors

\begin{eqnarray}
\widetilde{Z}^{\alpha} &=& \frac{(1+N)}{|1+N|}\:\ast \xi^{\alpha\lambda}\:\ast A_{\lambda} \:
/ \: (\:\sqrt{\ast A_{\mu} \ \ast \xi^{\mu\sigma} \
\ast \xi_{\nu\sigma} \ \ast A^{\nu}}\:) = \frac{(1+N)}{|1+N|}\:Z^{\alpha} \ , \label{ZTR}\\
\widetilde{W}^{\alpha} &=& \frac{(1+N)}{|1+N|}\:\ast \xi^{\alpha\lambda}\:\ast \xi_{\rho\lambda}\:\ast A^{\rho} \:
/ \: \: \sqrt{-Q/2} \: \sqrt{\ast A_{\mu} \ \ast \xi^{\mu\sigma} \
\ast \xi_{\nu\sigma} \ \ast A^{\nu}}\: = \frac{(1+N)}{|1+N|}\:W^{\alpha} \ .\label{WTR}
\end{eqnarray}

Therefore, from equations (\ref{ZTR}-\ref{WTR}) we infer that there are two possibilities, either we have $1+N>0$ or $1+N<0$. If $1+N<0$ we would obtain a local tetrad transformation represented by the full inversion or minus the identity two by two which is a valid case outside the Kernel. If the case is that $1+N>0$ then the local tetrad transformation on plane two would truly be in the Kernel fulfilling equations (\ref{COEFFNKER}-\ref{COEFFMKER}).  This is a case where the projection of equation (\ref{ASUM4}) on plane two inside the vectors (\ref{V3}-\ref{V4}) would result in $\Lambda^{\mu}=N\:A^{\mu}$ and we know that the gradient of $\Lambda$ in equation (\ref{ASUM4}) satisfies six integrability conditions. It is as if we chose the local gauge transformation to be $\Lambda^{\mu}=N\:A^{\mu}$. Immediately, we notice from equations (\ref{V3PUREGAUGE2}-\ref{V4PUREGAUGE2}) and (\ref{ZPUREGAUGE}-\ref{WPUREGAUGE}) that this kind of local gauge transformations are associated to projections of vector potentials that are related to the gradient of the local gauge transformation scalar through a local scalar factor which happens to be $\frac{1}{N}$ as in equation (\ref{ASUM3}). These would be gauge transformations associated to a potential that is pure gauge by itself and we are not considering these cases.

Conditions (\ref{COEFFNKER}-\ref{COEFFMKER}) would lead to exactly an analogous result as for plane one but this time on the local plane two.

\subsection{Kernel for the mapping fusion for planes one and two}
\label{fusionkernel}

However we can go one step forward and integrate the study of the Kernel on the local plane one and its orthogonal plane two in just one single package. We are mapping simultaneously the same $\Lambda^{\mu}$ into both orthogonal planes tetrad transformations. We do this by considering non-trivial the choice $X_{\alpha} = Y_{\alpha} = A_{\alpha} \mapsto  A_{\alpha} + \Lambda_{,\alpha}$. We assume that cases like the Reissner-Nordstr\"{o}m for which this choice of gauge vector in plane two would be ill-advised can be studied separately. To summarize this last stretch of our study, we are using the same potential to gauge $X_{\alpha} = A_{\alpha}$ and $Y_{\alpha} = A_{\alpha}$. We are also using simultaneously the same local scalar gradient in order to produce a local tetrad gauge transformation in both plane one and plane two in order to study their Kernels simultaneously. By putting together equations (\ref{ASUM2}) and (\ref{ASUM4}) we summarize the expression of the gradient of the local scalar $\Lambda$ in both Kernels with,

\begin{eqnarray}
\Lambda^{\mu} =  -C\:(A_{\alpha}\:U^{\alpha})\:U^{\mu} + N\:(A_{\alpha}\:W^{\alpha})\:W^{\mu}    \ .\label{ASUM6}
\end{eqnarray}

It is not difficult to prove that equation (\ref{ASUM6}) is a special case of the following general equation,

\begin{eqnarray}
(-Q/2)\:\Lambda^{\alpha} = -C \: V_{(1)}^{\alpha} - D \: V_{(2)}^{\alpha}
+ M \: V_{(3)}^{\alpha} + N \: V_{(4)}^{\alpha} \ . \label{Lambdainv}
\end{eqnarray}

For the simultaneous Kernel $D=M=0$. When we invert equation (\ref{ASUM6}) we obtain,

\begin{eqnarray}
A^{\mu} =  -\frac{1}{C}\:(\Lambda_{\alpha}\:U^{\alpha})\:U^{\mu} + \frac{1}{N}\:(\Lambda_{\alpha}\:W^{\alpha})\:W^{\mu}    \ .\label{ASUM7}
\end{eqnarray}

We have also used equations (\ref{ORTHOKER}) and (\ref{ORTHOKER2}). Now, in this fusion of both Kernels, N and M are given by,

\begin{eqnarray}
N + 1 &=&(-Q/2)\:V_{(4)\sigma}\:\Lambda^{\sigma} / (\:V_{(3)\beta}\:
V_{(3)}^{\beta}\:) + 1 > 0 \ ,\label{COEFFNKERL}\\
M &=&(-Q/2)\:V_{(3)\sigma}\:\Lambda^{\sigma} / (\:V_{(4)\beta}\:
V_{(4)}^{\beta}\:) = 0 \ .\label{COEFFMKERL}
\end{eqnarray}

We notice that in equations (\ref{COEFFNKERL}-\ref{COEFFMKERL}) we are no longer considering $\ast \Lambda^{\sigma}$ but $\Lambda^{\sigma}$ as in plane one, as can be compared with equations (\ref{COEFFNKER}-\ref{COEFFMKER}). In equation (\ref{ASUM6}) the potential vector $A^{\mu}$, and the tetrad vectors $U^{\mu}$ and $W^{\mu}$ are supposed to be known. We would end up with a Kernel associated to two local scalars C and N and eight derivatives of C and N. We would also have six integrability conditions on $\Lambda^{\mu}$ plus the two conditions (\ref{COEFFCKER}) and (\ref{COEFFNKER}). Again we would find that for the Kernel both projections of the potential on both local orthogonal planes would be proportional to the same gradient of the scalar $\Lambda^{\mu}$.

We conclude this section by saying that as long as our choice either on the local plane one or two for the gauge vectors $X^{\rho}$ and $Y^{\rho}$ is not the trivial pure gauge $X^{\rho} = \Lambda^{\rho}$ and $Y^{\rho} = \ast \Lambda^{\rho}$, or a scalar factor of the local scalar gradients as in equations (\ref{ASUM}-\ref{ASUM2}), (\ref{ASUM3}-\ref{ASUM4}) or (\ref{ASUM6}-\ref{ASUM7}), then, for the mappings given in reference \cite{A} by equations (58-59) for boosts or (91-92) for spatial rotations, also given by equations (\ref{TN1N}-\ref{TN2N}) and (\ref{TN3}-\ref{TN4}) in this paper, the only elements in the Kernel will be the constant gauge scalars $\Lambda = constant = \ast \Lambda$. Constant scalars with $\Lambda^{\mu} = 0 = \ast \Lambda^{\mu}$. Making both these transformations morphisms with trivial Kernel. In order to summarize all the results in this section we state the following,

\newtheorem {guesslb2}[guesslb1] {Theorem}
\begin{guesslb1}
The Kernel of the mapping between the local group of electromagnetic gauge transformations and the local group of tetrad transformations in the proper sector on the local blade one, corresponding to equations (58-59) in manuscript \cite{A} and that is, the boosts, also given by equations (\ref{TN1N}-\ref{TN2N}) in this paper, and boosts composed with the full inversion, corresponding to equations (60-61) in manuscript \cite{A}, is just composed by the group $\mathrm{PGB2}$ as long as the choice for tetrad gauge vector is not pure gauge $X^{\mu} \neq \Lambda^{\mu}$ or the pure gauge multiplied by a local scalar $X^{\mu} \neq \frac{1}{C}\:\Lambda^{\mu}$ with $1+C>0$ which is equivalent to pure gauge for the skeleton-gauge vector tetrad structure on plane one. The group $\mathrm{PGB2}$ is of measure zero. It is just gradients of scalars in a local plane in a four-dimensional spacetime. There will be an isomorphism between the group of local electromagnetic gauge transformations minus the set $\mathrm{PGB2}$ and $\mathrm{LB1}$.
\end{guesslb1}

\begin{guesslb2}
The Kernel of the mapping between the local group of electromagnetic gauge transformations and the local group of tetrad transformations on the local blade two, corresponding to equations (91-92) in manuscript \cite{A} and that is, spatial rotations, also given by equations (\ref{TN3}-\ref{TN4}) in this paper, is just composed by the group $\mathrm{PGB1}$ as long as the choice for tetrad gauge vector is not pure gauge $Y^{\mu} \neq \ast \Lambda^{\mu}$ or the pure gauge multiplied by a local scalar $X^{\mu} \neq \frac{1}{N}\:\ast \Lambda^{\mu}$ with $1+N>0$ which is equivalent to pure gauge for the skeleton-gauge vector tetrad structure on plane two. The group $\mathrm{PGB1}$ is of measure zero. It is just gradients of scalars in a local plane in a four-dimensional spacetime. There will be an isomorphism between the group of local electromagnetic gauge transformations minus the set $\mathrm{PGB1}$ and $\mathrm{LB2}$.
\end{guesslb2}

\begin{guesslb2}
The mapping between $\mathrm{U}(1)$ and $\mathrm{LB1} \otimes \mathrm{LB2}$ will be an isomorphism. The Kernel of this mapping will be just constant gauge transformations.
\end{guesslb2}

The mapping between $\mathrm{U}(1)$ and $\mathrm{LB1} \otimes \mathrm{LB2}$ will be an isomorphism. We can agree with the referee in the general sense that this isomorphism would be piecewise. Because we have in the local plane one boosts which are hyperbolic rotations, boosts composed with full inversions which are the composition of two reflections, boosts composed with spacetime reflections and boosts composed with full inversions and spacetime reflections. In the local plane two we have spatial rotations. It is in a general sense a piecewise isomorphism.

\subsection{Injectivity and surjectivity of the mapping}
\label{injsur}

We can ask if it is possible to map two different scalar functions $\Lambda_{1}$ and $\Lambda_{2}$ into the same Lorentz transformation both in blade one (not including the group $\mathrm{PGB2}$) or both in blade two (not including the group $\mathrm{PGB1}$) at the same point in spacetime. If this is possible, then, we can first generate a Lorentz transformation by $\Lambda_{1}$ and then another one by $-\Lambda_{2}$. The result should be the identity, because $-\Lambda_{2}$ generates the inverse Lorentz transformation of $\Lambda_{1}$. Therefore $\Lambda_{1}-\Lambda_{2}$ must be a constant. For the fusion case in section \ref{fusionkernel} the analysis would be similar. Summarizing, the injectivity remains proved, for more details about this proof see reference \cite{A}. The last point to make clear is related to the image of this mapping. The question to answer is if the image of this mapping is a subgroup of LB1. By subgroup we mean the following. LB2 is just $\mathrm{SO}(2)$, we are not concerned with the local plane two. In the local plane one LB1 is made up of $\mathrm{SO}(1,1)$ which is the subgroup of boosts plus the full discrete inversion or minus the identity and the spacetime reflection which is a discrete transformation given by $\Lambda^{o}_{\:\:o} = 0$, $\Lambda^{o}_{\:\:1} = 1$, $\Lambda^{1}_{\:\:o} = 1$, $\Lambda^{1}_{\:\:1} = 0$. The boosts plus the full inversion could be deemed as a subgroup of LB1 or even the boosts plus the reflection. The question is if the image of the mapping equals the codomain LB1 or just a subgroup. Let us develop the proof that the image coincides with the codomain LB1 in steps. Let us suppose first that there is a certain local gauge transformation $\Lambda$, such that $1+C > 0$, with $-1 < C < 0$. Then, there is always the transformation $n \: \Lambda$ with $n$ a natural number, $C_{n} = n\:C$. $1+C_{n} = 1+n\:C = 1 - n\: \mid C \mid$. For $n$ sufficiently large $1+C_{n}$ is going to become negative. We start with a boost and by multiplying the $\Lambda$ by a suitable $n$ we end up with a boost composed with a full inversion. If $C > 0$, there is the transformation $-n \: \Lambda$ with $n$ a natural number, $C_{n} = -n\:C$. $1+C_{n} = 1-n\:C = 1 - n\: \mid C \mid$. Once more for $n$ sufficiently large $1+C_{n}$ is going to become negative. Following similar ideas, but now if $1+C > D > 0$ we can prove for instance, that if $-1 < C < 0$, for $n$ sufficiently large, $D_{n} > 0 > 1+C_{n}$, and analogous for $C > 0$. Let us proceed to our particular case where we consider the following situation at some point in spacetime. $1 + C > D > 0$, where $0 > C > -1$. This last case represents a proper tetrad transformation, a local boost. Next, let us consider a new gauge transformation at the same point, $A\:\Lambda$, where $A$ is a constant. Then, the new $C_{new} = A\: C$ and $D_{new} = A\: D$ characterize the new tetrad transformation on blade one. In addition let this constant $A$ be such that $1 - A\:\mid C \mid = \epsilon$, $0 < \epsilon \ll 1$. The idea behind all this introduction would be to turn the original boost tetrad transformation into an improper transformation. For instance an improper transformation that satisfies $A\:D = D_{new} > 1 + C_{new} = 1 - A\:\mid C \mid = \epsilon$. This last improper condition is equivalent, after some algebra, to demanding that $\epsilon < 1 / (1 + \mid C \mid / D)$. Knowing that $A = (1 - \epsilon) / \mid C \mid$, by choosing at the point under consideration a constant $\epsilon$ that satisfies the inequality given above, $\epsilon < 1 / (1 + \mid C \mid / D)$, we would be generating an improper tetrad transformation , $A\:\Lambda$, out of a proper boost $\Lambda$. We are getting a boost composed with the reflection. The meaning of this result is nothing but stating that the image of the local group of electromagnetic gauge transformations into the local group of tetrad transformations LB1 is not a subgroup of the $\mathrm{LB1}$ group, for more details about this proof see reference \cite{ASU3}. Simply because we proved that a local electromagnetic gauge transformation $\Lambda$ inducing a proper local Lorentz tetrad transformation, can be turned into an improper tetrad Lorentz transformation through the new local electromagnetic gauge transformation $A\:\Lambda$. This last remark is equivalent to say that this mapping developed in paper \cite{A} is surjective. The image is LB1, and not one of its subgroups.



\end{document}